\documentclass[11pt,twoside]{article}
\usepackage{geometry}
\usepackage{graphicx}
\usepackage[english]{babel}
\pdfoutput=1
\usepackage[utf8]{inputenc}
\usepackage{hyperref}
\usepackage{amsfonts}
\usepackage{interval}
\usepackage{esvect}
\usepackage{comment}
\usepackage{amsthm,amsmath,amssymb,amsfonts,nccmath,mathtools,textcomp,graphicx,makeidx,color,gensymb,mhchem,url}
\usepackage{natbib,cite}
\usepackage{indentfirst}
\usepackage{enumitem}
\usepackage{multirow}
\usepackage{float}
\usepackage{cancel}
\usepackage{algorithm}
\usepackage[table]{xcolor}
\usepackage{longtable}
\usepackage{pgf, tikz}
\usepackage{tikz-network}
\usetikzlibrary{positioning, shapes, shadows, arrows, snakes,patterns}
\usetikzlibrary{arrows, automata}
\usepackage{pgfplots}

\tikzset{node_style/.style={draw,circle,line width=.1mm, inner sep=0,font=\fontsize{7}{10}\selectfont}}
\tikzset{edge_style/.style={draw=black, ultra thick, line width=.1mm,font=\fontsize{7}{10}\selectfont}}

\usepackage{array,multirow}

\usepackage{array}
\newcolumntype{C}[1]{>{\centering\arraybackslash}m{#1}}

\usepackage[flushleft]{threeparttable}

\usepackage{pdflscape}

\makeatletter
\newcommand{\thickhline}{%
    \noalign {\ifnum 0=`}\fi \hrule height 1pt
    \futurelet \reserved@a \@xhline
}
\newcolumntype{"}{@{\hskip\tabcolsep\vrule width 1pt\hskip\tabcolsep}}
\makeatother

\newcolumntype{P}[1]{>{\centering\arraybackslash}p{#1}}
\newcolumntype{M}[1]{>{\centering\arraybackslash}m{#1}}

\usepackage{array}
\newcolumntype{?}{!{\vrule width 1pt}}

\usepackage{makecell}

\usepackage{esvect}

\usepackage{listings}

\usepackage{hhline}  
\usepackage[table]{xcolor}

\def\titleab#1{{\Large\bf  \begin{flushleft} #1 \vspace{0pt} \end{flushleft}}}
\def\authors#1{{ \bf \begin{flushleft} #1 \vspace{0pt} \end{flushleft}}}
\def\university#1{{ \begin{flushleft} #1 \vspace{0pt} \end{flushleft}}}
\def\inst#1{\unskip$^{#1}$}

\newcommand{\keywords}[1]{\bigskip \noindent {\bf Keywords:} \ #1}

\newtheorem{definition}{Definition}[section]
\newtheorem{example}{Example}[section]
\newtheorem{note}{Note}[section]

\usepackage[font=scriptsize]{caption} 
\usepackage[font=scriptsize]{subcaption}

\usepackage{adjustbox}

\usepackage{array}
\newcolumntype{C}[1]{>{\centering\arraybackslash}m{#1}}

\usepackage[flushleft]{threeparttable}
\usepackage[customcolors]{hf-tikz}
\hfsetfillcolor{gray!10}
\hfsetbordercolor{gray}

\newcommand\blfootnote[1]{%
  \begingroup
  \renewcommand\thefootnote{}\footnote{#1}%
  \addtocounter{footnote}{-1}%
  \endgroup
}

\begin{document}
%
%
\titleab{Centrality Measures in Interval-Weighted Networks}

\authors{
  H\'elder Alves\textsuperscript{*}\inst{1}\blfootnote{* Corresponding author: H\'elder Alves, helder.alves@isssp.pt, Porto, Portugal}
  Paula Brito\inst{2},
  Pedro Campos\inst{2}
}

\university{
  \inst{1} ISSSP, Porto Institute of Social Work \& LIAAD INESC TEC, Portugal, \href{mailto:author1@mail.pt}{helder.alves@isssp.pt}\\
  \inst{2} FEP, University of Porto \& LIAAD INESC TEC, Portugal, \href{mailto:author2@mail.pt}{mpbrito@fep.up.pt}\\
}


\noindent\hrulefill
\begin{abstract}
\noindent
Centrality measures are used in network science to evaluate the centrality of vertices or the position they occupy in a network. There are a large number of centrality measures according to some criterion. However, the generalizations of the most well-known centrality measures for weighted networks, degree centrality, closeness centrality, and betweenness centrality have solely assumed the edge weights to be constants. This paper proposes a methodology to generalize degree, closeness and betweenness centralities taking into account the variability of edge weights in the form of closed intervals (Interval-Weighted Networks -- IWN). We apply our centrality measures approach to two real-world IWN. The first is a commuter network in mainland Portugal, between the 23 NUTS 3 Regions. The second focuses on annual merchandise trade between 28 European countries, from 2003 to 2015.\\


\end{abstract}

\noindent\hrulefill
\section{Introduction}

The study of the \textit{centrality measures} is one of the most important topics in network science~\citep{Borgatti:2005je,Brandes:2008gb,Lu:2016bv,Barabasi:2016vs,Brandes:2016id,Ghalmane:2019ev}. One of the questions that naturally arise when analysing a network is: \textit{``Which are the central vertices in the network?''}~\citep{Newman:2018ur}. The answer to this question depends on what we mean by \textit{important}. Even though there is no general consensus on the exact definition of ``importance'', in a \textit{structural} approach, which is the most common, the importance of a vertex is usually related to the concept of being the most connected vertex or being positioned in the center of the network~\citep{Freeman:1977vn,Freeman:1979wx,Bonacich:1987up,Borgatti:2006cf}. Essentially, a vertex positioned in the center of a network has advantages over other vertices, as it is directly linked to many other vertices (has more edges) or acts as an intermediary in communicating with other vertices, either at speed (it is closer) or in the flow control with which it reaches the other vertices (it is between). Identifying these ``vital'' vertices allow us to control the outbreak of epidemics, to conduct advertisements for e-commercial products, to predict popular scientific publications, and so on~\citep{Lu:2016bv}. There are a large number of centrality measures that capture the varying importance of the vertices (vertex-level measures) in a network according to some criterion, such as reachability, influence, embeddedness, control the flow of information~\citep{Rodrigues:2019ff}. Some of these most well known measures are \textit{degree centrality} and \textit{closeness centrality}~\citep{Sabidussi:1966wp,Freeman:1979wx}, \textit{Betweeness centrality}~\citep{Freeman:1977vn}, and \textit{Eigenvector centrality}~\citep{Bonacich:1972dt} along with its variants~\citep{Bonacich:1987up} and \textit{Page rank}~\citep{Brin:1998vm}. Other centrality measures are \textit{Katz centrality}~\citep{Katz:1953un}, \textit{Information centrality (or S-Z centrality index)}~\citep{Stephenson:1989ug}, \textit{Betweeness centrality based on flow networks}~\citep{Freeman:1991un}, \citet{Valente:1998vp} \textit{integration and radiality measures}, \textit{Centrality based on game theory}~\citep{Gomez:2003ud}, \textit{Betweeness centrality based on random walks}~\citep{Newman:2005vv}, among others.

Recently, \citet{Gomez:2013ee} introduced a centrality measure based on \textit{bi-criteria network flow}. \citet{Martin:2014tr} proposed a new centrality measure based on the leading eigenvector of the Hashimoto or nonbacktracking matrix. \citet{Du:2014hp} presented TOPIS as a new measure of centrality. \citet{Lu:2016kl} suggested a novel measure of node influence based on comprehensive use of the degree method, H-index and coreness metrics. \citet{Brandes:2016id} propose a variant notion of distance that maintains the duality of closeness-as-independence with betweenness also on valued relations. \citet{Qiao:2017js} introduced a novel \textit{entropy centrality} approach. \citet{Wu:2019mi} introduced eigenvector multicentrality based in a tensor-based framework. \citet{Ghalmane:2019ev} extended all the standard centrality measures defined for networks with no community structure to modular networks \textit{Modular centrality}. \citet{Zhang:2020gi} derive a new centrality index \textit{resilience centrality}. A comprehensive explanation of some of these measures can be found in~\citet{Lu:2016bv} and the book by~\citet{Newman:2018ur}.\\

In this paper, we focus only on the most influential and well-known centrality measures, degree, closeness and betweenness~\citet{Freeman:1979wx}. Initially these three measures were formalized for \textit{binary} (unweighted) networks. However, as~\citet{Freeman:1979wx} refers, binary representations fail to capture any of the important variability in strength, and naturally these measures were later extended to \textit{weighted} networks. Firstly, by allowing to capture the strength of an edge focusing only on edge weights~\citep{Newman:2001kc,Brandes:2001wm,2004PNAS..101.3747B}. Secondly by taking into consideration both the weight and the number of edges including a \textit{tuning parameter}, $\alpha$~\citep{Opsahl:2010in}\footnote{\citet{Opsahl:2010in} point out some caveats of these generalizations: first, the edge weight must have a ratio scale, otherwise the mean weight has no real meaning; and secondly, it is difficult to determine the most appropriate value of the tuning parameter $\alpha$.}. Moreover, as some centrality measures (closeness and betweenness) are based on the shortest paths, they do not take into account the flow of the edge content along non-shortest paths. Thus,~\citet{Freeman:1991un} proposed a betweenness measure based on Ford and Fulkerson's (FF) model of \textit{network flows} (Ford and Fulkerson,~\citeyear{Ford:1956vc}, \citeyear{Ford:1957vq}, \citeyear{Ford:26m8xm4j}), thereby allowing to account for the flow of the edges of the entire network.\\

\begin{figure}[ht]
	\centering
    	\includegraphics[scale=.6, clip, trim={0cm 11.25cm 0cm 11.25cm}]{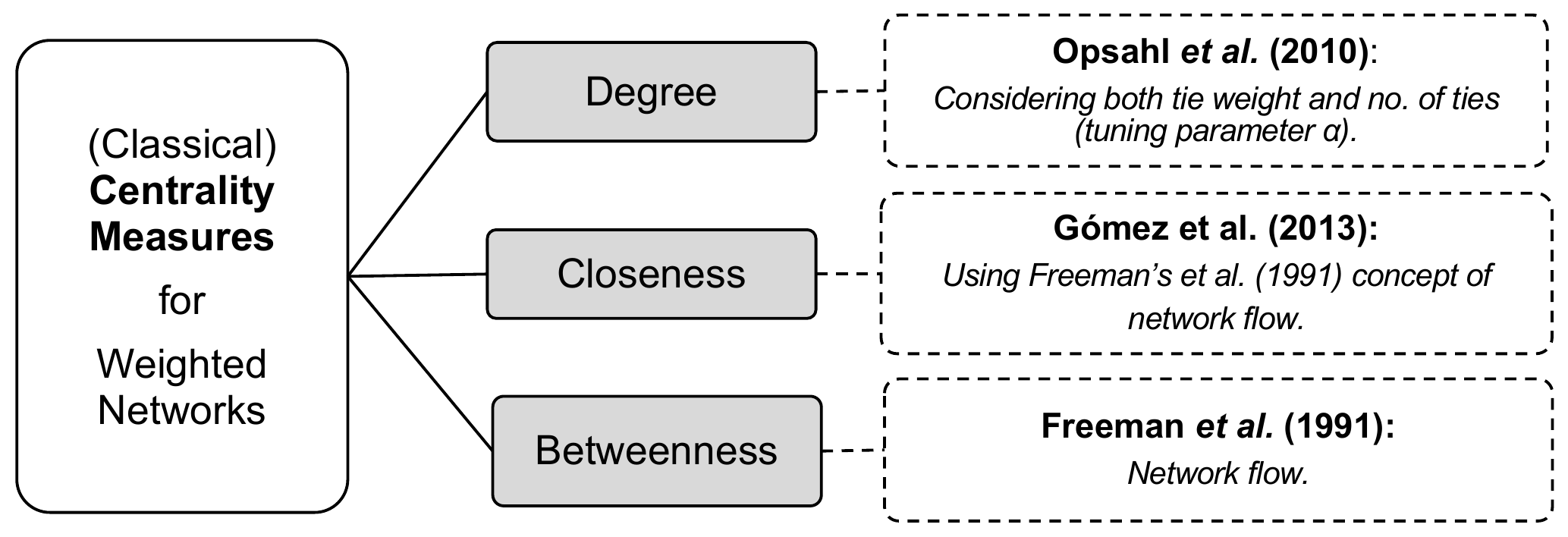}
	\caption{Scheme of the generalizations made for interval-weighted networks (IWN) of the three (classical) measures of vertex centrality, \textit{degree}, \textit{closeness} and \textit{betweenness}. In dashed lines are indicated the approaches adopted when generalizing to IWN.}
	\label{chp6_fig:esquema_medidas_centralidade}
\end{figure}

Nevertheless, none of the above methodologies allows accounting for the variability observed in the original data. The main contribution of this paper is the development of three new measures for degree, closeness and betweenness, taking into account the networks' variability of edge weights in the form of closed intervals. This way, a closed interval may be used to model the precise information of an objective entity that comprehends intrinsic variability \textit{(ontic view)}, i.e., an interval $A$ is a value of a set-valued variable $X$, so we can write $X=A$~\citep{Couso:2014du,Grzegorzewski:2016fq}. We call such networks \textit{interval-weighted networks} (IWN) (see Figure~\ref{chp6_fig:conversion_to_direct_net}), and consequently, we name these measures the \textit{interval-weighted degree (IWD)}, \textit{interval-weighted flow betweenness (IWFB)} and \textit{interval-weighted flow closeness (IWFC)}. Our methodology is depicted in Figure~\ref{chp6_fig:esquema_medidas_centralidade}. The dashed lines indicate the methods followed in this paper in the generalization of the centrality measures for interval-weighted networks.\\

The remaining of the paper is organized as follows. We start by briefly introducing the basic terms and concepts of interval arithmetic and interval order relations and we propose a new order relation between intervals (Section~\ref{Interval_Analysis}). Then, in Section~\ref{chp6_Sec_centrality_weighted_networks} we recall the centrality measures for weighted networks. First, we present the degree for the general case and also taking into account both edge weights and the number of edges introducing a tuning parameter $(\alpha)$. Second, we define the concept of flow networks for the case of an (undirected) interval-weighted network and corresponding centrality measures, flow closeness and flow betweenness. In Section~\ref{chp6_Sec_centrality_measures_IWN}, we generalize degree, flow betweenness and flow closeness to the case of interval-weighted networks. In Section~\ref{applications}, we apply our generalizations of degree, flow betweenness and flow closeness to the case of interval-weighted networks in two real-world applications. Finally, in Section~\ref{conclusion} we conclude and discuss the outcomes obtained with our methodology.

\section{Interval Analysis}
\label{Interval_Analysis}

Let $\underline{x},\overline{x} \in \mathbb{R}$ such that $\underline{x} \leqslant \overline{x}$. An interval number $[\underline{x},\overline{x}]$ is a closed bounded nonempty real interval, given by
$
[\underline{x},\overline{x}]=\{x \in \mathbb{R}\colon \underline{x} \leqslant x \leqslant \overline{x}\}
$,
where $\underline{x}=\min([\underline{x},\overline{x}])$ and $\overline{x}=\max([\underline{x},\overline{x}])$ are called, respectively, the \textup{lower} and \textup{upper} bounds (endpoints) of $[\underline{x},\overline{x}]$.
The set $[\mathbb{R}]$ of interval numbers is a subset of the powerset of $\mathbb{R}$ such that 
$
[\mathbb{R}]=\big\{X \in \wp\, (\mathbb{R})\colon (\exists \underline{x} \in \mathbb{R})\, (\exists \overline{x} \in \mathbb{R})\, (X=[\underline{x},\overline{x}])\big\}.
$
Since, corresponding to each pair of real constants $\underline{x},\overline{x}\ (\underline{x}\leqslant\overline{x})$ there exists a closed interval $[\underline{x},\overline{x}]$, the set of interval numbers is \textit{infinite}. We say that $X$ is \textup{degenerate} if $\underline{x}=\overline{x}$. By convention, a degenerate interval $[x,x]$ is identified with the real number $x$ (e.g., $1=[1,1]$).
For any two intervals $X=[\underline{x}, \overline{x}]$ and $Y=[\underline{y}, \overline{y}]$, in terms of the intervals' endpoints, the four classical operations of real arithmetic can be extended to intervals as follows~\citep{Moore:2009uc}:
\begin{itemize}
\setlength\itemsep{0.5pt} 
\item 
Interval addition, 
$
X+Y=[\underline{x}, \overline{x}] + [\underline{y}, \overline{y}] = [\underline{x}+\underline{y}, \overline{x}+\overline{y}]
$;
\item Interval multiplication,
$
X\cdot Y=[\underline{x}, \overline{x}] \cdot [\underline{y}, \overline{y}]=\big\lbrack\min\{\underline{x}\underline{y},\underline{x}\overline{y},\overline{x}\underline{y},\overline{x}\overline{y}\}, \max\{\underline{x}\underline{y},\underline{x}\overline{y},\overline{x}\underline{y},\overline{x}\overline{y}\}\big\rbrack
$;
\item Interval subtraction,
$
X-Y=X+(-Y)
$
where $-Y=[-\overline y,-\underline y]$ (reversal of endpoints)\footnote{It should be noted that the subtraction of two equal intervals is not $\interval{0}{0}$ (except for degenerate intervals). This is because $X-X=\{x-y\colon x \in X,\, y \in Y \}$, rather than $\{x-x\colon x \in X\}$~\citep{Moore:2009uc}. For example, $\interval{1}{2}-\interval{1}{2}=\interval{-1}{1}$.}.
\item 
Interval division for any $X\in\mathbb{R}$ and any $Y\in\lbrack\mathbb{R}\rbrack_{\widetilde{0}}$, is defined by
$
X \div Y = X\cdot(Y^{-1})
$,
where $Y^{-1}=1/Y=[1/ \overline y,1/ \underline y]$, assuming that $0 \not\in Y$.
\end{itemize}

Intervals can also be represented by their \textit{midpoint (or mean, or center)} $m$ and \textit{half-width} (or radius), $rad$. So, $X=\interval{\underline{x}}{\overline{x}}=\langle m(X),rad(X) \rangle$, where $m(X)=\mfrac{(\underline{x} + \overline{x})}{2}$ and $rad(X)=\mfrac{(\overline{x} - \underline{x})}{2}$.

The \textit{infimum} between two intervals $X=\interval{\underline{x}}{\overline{x}}$ and $Y=\interval{\underline{y}}{\overline{y}}$ is defined to be $
\inf(X,Y)=\lbrack\inf(\underline{x},\underline{y}),\inf(\overline{x},\overline{y})\rbrack$. Similarly, the \textit{supremum} between two intervals $X=\interval{\underline{x}}{\overline{x}}$ and $Y=\interval{\underline{y}}{\overline{y}}$ is defined to be $
\sup(X,Y)=\lbrack\sup(\underline{x},\underline{y}),\sup(\overline{x},\overline{y})\rbrack$~\citep{Dawood:2011vh}.
An operation whose operands are intervals $(\interval{\underline{x}}{\overline{x}})$, and whose result is a point interval (or a real number) is called a \textit{point interval operation}, such as the: \textit{infimum} $\inf(\interval{\underline{x}}{\overline{x}})=\min(\interval{\underline{x}}{\overline{x}})=\underline{x}$ and \textit{supremum} $\sup(\interval{\underline{x}}{\overline{x}})=\max(\interval{\underline{x}}{\overline{x}})=\overline{x}$. Finally, another important definition of a point interval operation is the Hausdorff distance (or metric) between two intervals~(see, e.g., \citealp{billard2006symbolic}):
$d(X,Y)=d(\interval{\underline{x}}{\overline{x}}, \interval{\underline{y}}{\overline{y}})=\max\{|\underline{x}-\underline{y}|,|\overline{x}-\overline{y}|\}$.

\paragraph{Interval arithmetic pitfalls:}
Useful properties of ordinary real arithmetic fail to hold in classical interval arithmetic. Some of the main disadvantages of the classical interval theory are~\citep{Dawood:2011vh}: (i) \textit{Interval dependency} -- subtraction and division are not the inverse operations of addition and multiplication, respectively; (ii) \textit{Distributive law does not hold} -- only a subdistributive law is valid -- $\forall X,Y,Z \in [\mathbb{R}]\ Z\times (X+Y) \subseteq Z\times X + Z \times Y$.


\subsection{Interval order relation}
\label{ordering_intervals}
Till date one main dilemma in using interval data for decision problems is perhaps the choice of an appropriate interval order relation. Unlike real numbers that are ordered by a strict transitive relation ``$<$'' (if $a<b$ and $b<c$, then $a<c$ for any $a,b$, and $c \in \mathbb{R}$), the ranking of intervals is \textit{not symmetric}, and as consequence, in many situations, the definitions cannot differentiate two intervals in general, even though they can be applied efficiently to solve the prescribed models~\citep{Karmakar:2014jo}. As a consequence, theoretically, intervals can only be \textit{partially ordered} in $[\mathbb{R}]$. According to \citet{Moore:2009uc}, two transitive order relations can be defined for intervals: (i) $X < Y \Leftrightarrow \overline{x} < \underline{y}$, and (ii) $X \subseteq Y  \Leftrightarrow \underline{y} \leqslant \underline{x}\ \text{and}\ \overline{x} \leqslant \overline{y}$ (set inclusion)\footnote{\textit{Set inclusion} ``$\subseteq$'' is a partial order between intervals, which is reflexive, antisymmetric and transitive.}. Nevertheless, when a choice has to be made among alternatives, the comparison is indeed needed. There are several different approaches in the literature for ordering intervals~\citep{Hu:2006tq,Sengupta:2009wk,Guerra:2012hg,Stefanini:2019ie}. A detailed description and comparison between these and other ranking definitions is given in~\citep{Karmakar:2012vm}.

Bearing in mind the above definitions, and according to Hossain's methodology~\citep{Hossain:2010vq}, we may define the following order relation:

\begin{definition}\label{chp3_def:our_ranking}
Given two intervals $X,Y\in \mathbb{R}$, $X\leqslant Y$, iff $m(X)\leqslant m(Y)$. Furthermore $X<Y$ iff $X\leqslant Y$ and $X\neq Y$\footnote{This order relation also applies to the case when intervals are \textit{completely overlapping}, but $m(X)\neq m(Y)$.}. In the case where the midpoints of $X$ and $Y$ coincide $m(X)=m(Y)$, the intervals $X$ and $Y$ are said to be equivalent $X \approx Y$. We propose the following order relation: $X\leqslant Y$ is determined by choosing the interval that captures the \textit{``maximum variability''} between the two intervals, i.e., the interval with the highest radius. For example, if $m(X)=m(Y)$ the decision for $X\leqslant Y$ implies that $rad(Y) > rad(X)$.
\end{definition}

To exemplify and illustrate the applicability of these definitions in a simple network, we built three scenarios for ranking a pair of intervals. The results are shown in Table~\ref{chp3_tab:ranking_interval-weighted_triplets}.

\begin{table}[ht]
\caption{Interval ordering for the order relation ``$\leqslant$'', to choose the \textit{``greater interval''}.}\label{chp3_tab:ranking_interval-weighted_triplets}
\footnotesize
\centering
\renewcommand{\arraystretch}{1.4}
\begin{adjustbox}{max width=\textwidth}
\begin{tabular}{c|c|c|c}
\hline
& \multicolumn{3}{c}{Interval relations} \\\cline{2-4}
\multirow{4}{*}{Order relation ``$\leqslant$''}
&
\begin{tikzpicture}[inner sep=0pt, minimum size=5.25mm, auto,
   	node_style/.style={draw,circle,line width=.2mm, font=\fontsize{8}{10}\selectfont},
   	edge_style/.style={draw=black, line width=.2mm}]
    \node[node_style] (v1) at (0,1) {$v_{1}$};
    \node[node_style] (v2) at (1.5,2.25) {$v_{2}$};
    \node[node_style] (v3) at (2,1) {$v_{3}$};
    \draw[edge_style]  (v1) edge node[above,sloped,pos=0.5,font=\fontsize{8}{10}\selectfont] {$[1,3]$} (v2);
    \draw[edge_style]  (v1) edge node[below,sloped,pos=0.5,font=\fontsize{8}{10}\selectfont] {$[4,6]$} (v3);
    \node[fit=(current bounding box),inner ysep=2mm,inner xsep=0]{};  
\end{tikzpicture}
&
\begin{tikzpicture}[inner sep=0pt, minimum size=5.25mm, auto,
   	node_style/.style={draw,circle,line width=.2mm, font=\fontsize{8}{10}\selectfont},
   	edge_style/.style={draw=black, line width=.2mm}]
    \node[node_style] (v1) at (0,1) {$v_{1}$};
    \node[node_style] (v2) at (1.5,2.25) {$v_{2}$};
    \node[node_style] (v3) at (2,1) {$v_{3}$};
    \draw[edge_style]  (v1) edge node[above,sloped,pos=0.5,font=\fontsize{8}{10}\selectfont] {$[1,4]$} (v2);
    \draw[edge_style]  (v1) edge node[below,sloped,pos=0.5,font=\fontsize{8}{10}\selectfont] {$[3,6]$} (v3);
    \node[fit=(current bounding box),inner ysep=2mm,inner xsep=0]{};  
\end{tikzpicture}
&
\begin{tikzpicture}[inner sep=0pt, minimum size=5.25mm, auto,
   	node_style/.style={draw,circle,line width=.2mm, font=\fontsize{8}{10}\selectfont},
   	edge_style/.style={draw=black, line width=.2mm}]
    \node[node_style] (v1) at (0,1) {$v_{1}$};
    \node[node_style] (v2) at (1.5,2.25) {$v_{2}$};
    \node[node_style] (v3) at (2,1) {$v_{3}$};
    \draw[edge_style]  (v1) edge node[above,sloped,pos=0.5,font=\fontsize{8}{10}\selectfont] {$[2,5]$} (v2);
    \draw[edge_style]  (v1) edge node[below,sloped,pos=0.5,font=\fontsize{8}{10}\selectfont] {$[1,6]$} (v3);
    \node[fit=(current bounding box),inner ysep=2mm,inner xsep=0]{};  
\end{tikzpicture}
\\
&
\begin{tikzpicture}
	\draw [-latex, thick]  (0.5,0) -- (4,0);
	\draw (1,0) [style={font=\footnotesize}] node[below=1pt] {$1$};
    \draw (2,0) [style={font=\footnotesize}] node[below=1pt] {$3$};
    \draw (2.5,0) [style={font=\footnotesize}] node[below=1pt] {$4$};
    \draw (3.5,0) [style={font=\footnotesize}] node[below=1pt] {$6$};     
    \foreach \x in {1,2,2.5,3.5}
      \draw (\x cm,3pt) -- (\x cm,-3pt);
	\draw [thick] (1,0) rectangle (2,0.5);
	\draw [thick,pattern=north east lines] (2.5,0) rectangle (3.5,1);
     \draw [thin,dashed] (1.5,0) -- (1.5,0.5);
     \draw [thin,dashed] (3,0) -- (3,1);
    \node at (1,0) [above=.55,style={font=\small}] {$X$};
	\node at (3.5,0) [above=1.05,style={font=\small}] {$Y$}; 
\end{tikzpicture}
&
\begin{tikzpicture}
	\draw [-latex, thick]  (0.5,0) -- (4,0);
	\draw (1,0) [style={font=\footnotesize}] node[below=1pt] {$1$};
    \draw (2.5,0) [style={font=\footnotesize}] node[below=1pt] {$4$};
    \draw (2,0) [style={font=\footnotesize}] node[below=1pt] {$3$};
    \draw (3.5,0) [style={font=\footnotesize}] node[below=1pt] {$6$};     
    \foreach \x in {1,2,2.5,3.5}
      \draw (\x cm,3pt) -- (\x cm,-3pt);
	\draw [thick] (1,0) rectangle (2.5,0.5);
	\draw [thick,pattern=north east lines] (2,0) rectangle (3.5,1);
     \draw [thin,dashed] (1.75,0) -- (1.75,0.5);
     \draw [thin,dashed] (2.75,0) -- (2.75,1);
    \node at (1,0) [above=.55,style={font=\small}] {$X$};
	\node at (3.5,0) [above=1.05,style={font=\small}] {$Y$}; 
\end{tikzpicture}
&
\begin{tikzpicture}
	\draw [-latex, thick]  (0.5,0) -- (4,0);
	\draw (1,0) [style={font=\footnotesize}] node[below=1pt] {$1$};
    \draw (3.5,0) [style={font=\footnotesize}] node[below=1pt] {$6$};
    \draw (1.5,0) [style={font=\footnotesize}] node[below=1pt] {$2$};
    \draw (3,0) [style={font=\footnotesize}] node[below=1pt] {$5$};     
    \foreach \x in {1,1.5,3,3.5}
      \draw (\x cm,3pt) -- (\x cm,-3pt);
	\draw [thick,pattern=north east lines] (1,0) rectangle (3.5,1);
	\draw [thick] (1.5,0) rectangle (3,0.5);
     \draw [thin,dashed] (2.25,0) -- (2.25,0.5);
     \draw [thin,dashed] (2.25,0) -- (2.25,1);
    \node at (3,0) [above=.55,style={font=\small}] {$X$};
	\node at (1,0) [above=1.05,style={font=\small}] {$Y$}; 
\end{tikzpicture}
\\\hline
{\em Type of interval} & Non-overlapping & Partially overlapping & \makecell{Completely overlapping:\\$m(X)=m(Y)$} \\\hline
$X \cap Y$ & $X \cap Y=\varnothing$ & $X \cap Y \neq \varnothing$ & $X \cap Y \neq \varnothing \land X \subseteq Y$ \\
$d_H(X,Y)$ & $\max(|{-3}|,|{-3}|)=3$ & $\max(|{-2}|,|{-2}|)=2$ & $\max(|{1}|,|{-1}|)=1$ \\
$\inf(X,Y)$ & $\min([1,3],[4,6])=[1,3]$ & $\min([1,4],[3,6])=[1,4]$ & $\min([2,5],[1,6])=[1,5]$ \\
$\sup(X,Y)$ & $\max([1,3],[4,6])=[4,6]$ & $\max([1,4],[3,6])=[3,6]$ & $\max([2,5],[1,6])=[2,6]$ \\
$\langle midpoint,\operatorname{half-width} \rangle$ & $X=\langle 2,1\rangle;Y=\langle 5,1\rangle$ & $X=\langle 2.5,1.5\rangle;Y=\langle 4.5,1.5\rangle$ & $X=\langle 3.5,1.5\rangle;Y=\langle 3.5,2.5\rangle$\\
\hline
{Decision = choose the \textit{greater} interval} & $choose\ Y$ & $choose\ Y$ & $choose\ Y$ \\
\hline
\end{tabular}
\end{adjustbox}
\end{table}


\section{Centrality measures for weighted networks: \textit{degree}; \textit{flow betweenness} and \textit{flow closeness}}
\label{chp6_Sec_centrality_weighted_networks}

\paragraph{Degree centrality} of a vertex $i$ for weighted networks (or the vertex \textit{strength}) is defined as the sum of weights attached to edges connected to vertex $i$~\citep{2004PNAS..101.3747B}\footnote{For weighted social networks, \citet{Granovetter:1973wj} refers that the weight of an edge is generally a function of duration, emotional intensity, intimacy, and exchange of services, while for non-social weighted networks, it usually represents the capacity or capability of the edge~\citep{Hu:2008vt}.}. Usually it is formalized as:
\begin{equation}
\label{chp6_weighted-degree_1}
s_i=C^W_D(i)=\sum^n_{j=1} w_{ij} \qquad s_i>0
\end{equation}
where $w_{ij}$ is the entry of the $i$th row and $j$th column of the weighted adjacency matrix $W$.\\
\citet{Opsahl:2010in} include a tuning parameter, $\alpha$, to address the relative importance of the number of edges compared to edge weights associated with a vertex $i$, this is formalized as:
\begin{equation}
\label{chp6_weighted-degree_1_opsahl}
C^{W\alpha}_D (i)=k_i {\left(\frac{s_i}{k_i}\right)}^\alpha=k^{(1-\alpha)}_i s^{\alpha}_i
\end{equation}
where $k_i$ is the number of vertices that a focal vertex $i$ is connected to.

Briefly, when $0<\alpha<1$, both vertex degree and strength will be taken into account, when $\alpha>1$, the measure would positively value edge strength and negatively value the number of edges, for $\alpha=0$, the measure is solely based on the number of edges, and for $\alpha=1$ the measure is based only on edge weights (see \citealt{Opsahl:2010in} for details).

\citet{Garas:2012em} exemplifies the situation described above, through economic or commercial networks where weights generally play an important role (usually representing the flow of capital or the flow of trade). In these cases, the focus is usually on the vertices with higher strength, usually the most important. Thus, in such networks the presence of vertices with a high degree and relatively small strength can influence the results obtained by methods that are based only on the degree.

\paragraph{Flows in undirected weighted networks}
The reason why we opted for using Freeman's \citep{Freeman:1991un} approach of \textit{flow networks} to generalize both \textit{betweenness}~\citep{Freeman:1979wx,Opsahl:2010in} and \textit{closeness}~\citep{Newman:2001kc,Brandes:2001wm,Opsahl:2010in} to weighted networks, was mainly because, first~\citet{Freeman:1991un} and later~\citet{Newman:2005vv}, \citet{Brandes:2005ug} and \citet{,Borgatti:2005je}, pointed out that closeness and betweenness centrality measures based on the shortest paths do not take into account the flow of the edge content (e.g., information) along non-shortest paths, assuming that the edge content only flows along the shortest possible paths~\citep{Borgatti:2005je}. Therefore, these measures are unlikely to characterize human communication, disease proliferation, etc.~\citep{Barbosa:2018ki}. To contour this, \citet{Freeman:1991un} proposed a betweenness measure based on Ford and Fulkerson's (FF) model of \textit{network flows} (Ford and Fulkerson,~\citeyear{Ford:1956vc}, \citeyear{Ford:1957vq}, \citeyear{Ford:26m8xm4j}).\\

\begin{note}\label{note_network_conversion_undirected-directed}
Since both the flow networks and the Ford and Fulkerson algorithm~\citep{Ford:26m8xm4j} have been defined for direct networks, it is first necessary to transform an undirect network \textit{into} a direct network; this is done as follows:
for an undirected and connected  network $G=(V,E)$, where $V\neq \emptyset$ is a finite set of vertices and $E$ is a set of pairs of vertices called edges $E\subseteq\big\{(i,j)\colon i,j \in V \big\}$, the extension of Ford and Fulkerson algorithm~\citep{Ford:26m8xm4j} is done by considering two direct edges $(i,j)$ and $(j,i)$ \big(hereinafter $\{i,j\}$\big), one in each direction, for each edge in the original network. Considering the two direct edges between a given pair of vertices, when an edge is used in a flow, the other edge cannot be used~\citep{Freeman:1991un,Schroeder:2004uv,Gomez:2013ee}.
\end{note}

\begin{definition}[Undirected Flow Network]
\label{chp6_Def_flow_network}
Given a connected undirected network $G=(V,E)$, and a pair of vertices $s$ (source), and $t$ (sink) $\in V$, let $f(i,j)$ be the \textit{flow} in the edge $(i,j)\in E$, and let the maximum allowable flow on that edge be $c(i,j)$, its \textit{capacity} $(c\colon E\rightarrow \mathbb{R^+})$. A \textit{flow} in $G$ is a function $f\colon V\times V\rightarrow\mathbb{R}$ that satisfies the following properties:

\begin{itemize}
\setlength\itemsep{0.5pt} 
\item \textit{capacity constraint} -- the resources used by a flow on an edge cannot be greater than the capacity of that edge: $0\leqslant f(i,j)\leqslant c\left(\{i,j\}\right),\ \forall i,j\in V$;
\item \textit{skew symmetry} -- the network flow from vertex $i$ to $j$ is the negative of the network flow in the reverse direction: $f(i,j)=-f(j,i),\ \forall i,j\in V$ (and thus, $f(i,i)=-f(i,i)=0$);
 \item \textit{flow conservation} -- the sum of all flows that enter in a vertex (negative flows) plus the sum of all flows that leave that vertex (positive flows) is null: $\sum_{j\in V} f(i,j)=0,\forall i\in V-\{s,t\}$.
\end{itemize}
Thus, the value of a flow is the sum of all outgoing flow $f(i,j)$ from the source $s$, defined as: $|f|=\sum_{j\in V} f(s,j)=\sum_{j\in V} f(j,t)$.
\end{definition}

However, we are interested in finding the \textit{overall flow} between pairs of vertices along all the paths that connect them. To find out the maximum allowable flow (hereafter, \textit{max-flow}) on a flow network from any source $i$ to any sink $j$, we will use the algorithm developed by~\citet{Ford:26m8xm4j}\footnote{\citet{Ford:26m8xm4j} proved that the maximum flow (max-flow) from $i$ to $j$ is exactly equal to that \textit{minimum cut} (min--cut) capacity. The min--cut capacity from $i$ to $j$ is the smallest capacity of any of the $i-j$ cut sets.}. This algorithm uses flow-augmenting paths to increase existing flows in the network, so that in each iteration the flow is greater. The basic idea behind the FF algorithm for undirected networks, is as follows:

\begin{definition}[Ford and Fulkerson algorithm]
\label{chp6_Def_FF_algorithm}
Given a connected network $G=(V,E)$, where $V$ is a set of vertices and $E$ is a set of edges between these vertices $(E\subseteq \{\{i,j\}: i,j\in V\})$ , for a flow $f$ between a pair of vertices $(i,j)$, the forward \textit{residual capacity} from $i$ to $j$ is denoted by $c_f(i,j)=c\left(\{i,j\}\right)-f(i,j)$, where $c\left(\{i,j\}\right)$ is the forward capacity of $(i,j)$ (the order of vertices connected by an edge does not matter, as well as the associated capacity)\footnote{The residual capacity is used by the algorithm to determine how much flow can pass through a pair of vertices, and is used in the definition of the so called \textit{residual network.}}.

The \textit{residual network} is basically an auxiliary network used by the algorithm, defined as: given a network $G=(V,E)$, let $f$ be a flow in $G$, the \textit{residual network} induced by $f$ is a network $G_f=(V, E_f)$, where $E_f=\{(i,j)\in V\times V\colon c_f(i,j)>0\}$. The \textit{residual network} is always directed, either for directed or undirected networks. The \textit{residual capacity} from $j$ to $i$, in the backward direction of the edge $(i,j)$, is defined as $c_f(j,i)=c\left(\{j,i\}\right)+f(i,j)$. That is, the residual capacity is the additional flow one can send on an edge, possibly by cancelling some flow in opposite direction.

Let $p$ be a \textit{path} from $s$ to $t$ that is allowed to transverse edges in either the forward or backward direction, the residual capacity $c_f(p)$ of a path $p$ is the minimum residual capacity of its edges, that is, $c_f(p)=\min\limits_{(i,j)\in p} c_f(i,j)$. If $c_f(p)>0$, then $p$ is called an \textit{augmented path}. A value of total flow can be increased by adding the minimum capacity on each forward edge and subtracting it from every backward edge in the augmented path.

By repeating this process of finding the augmented paths on a flow network, the total flow can be increased to the maximum within capacity constraints.
\end{definition}

The pseudo-code bellow (see Algorithm~\ref{chp6_Alg:Ford&Fulkerson}) describes the Ford \& Fulkerson algorithm for undirected networks~\citep{Schroeder:2004uv}.
\pagebreak

\begin{algorithm}
\linespread{0.5}\selectfont
\caption{Pseudo-code: Ford and Fulkerson algorithm for undirected networks}
\label{chp6_Alg:Ford&Fulkerson}
\textbf{Input:} A connected undirected flow network $G=(V,E)$\\
\textbf{Output:} The maximum flow $f$ on $G$ 
\begin{enumerate}[label*=\arabic*:]
\setlength\itemsep{0cm}    
\item \textbf{for} each edge $\{i,j\}\in E$ \textbf{do}
\item $\qquad f[i,j]\leftarrow 0$
\item $\qquad f[j,i]\leftarrow 0$
\item \textbf{while} there exists a path $p$ from $s$ to $t$ with no cycles in the residual network $G_f$ \textbf{do}
\item $\qquad \Delta \leftarrow\min\{c_f(i,j)\colon(i,j)\in p\}$
\item $\qquad$ \textbf{for} each edge $(i,j)$ in $p$ \textbf{do}
\item $\qquad\qquad f[i,j]\leftarrow f[i,j]+\Delta$
\item $\qquad\qquad f[j,i]\leftarrow -f[i,j]$
\item \textbf{return} $f$
\end{enumerate}
\end{algorithm}

Having defined a flow network (Definition~\ref{chp6_Def_flow_network}) and the FF max-flow algorithm (Definition~\ref{chp6_Def_FF_algorithm}), next we present the respective \textit{flow centrality measures}.\\

\begin{note}
All the generalizations take into account only undirected and connected weighted networks, $G=(V,E)$, where $V\neq \emptyset$ is a finite set of vertices and $E$ is a set of positive weighted edges, where the weights measure the strength $w_{ij}\geqslant 0$, $E\subseteq\big\{(i,j)\colon i,j \in V \big\}$. For unweighted networks, we define $w_{ij}=1$ if there is an edge between vertices $i$ and $j$ and zero otherwise.
 \end{note}


\textit{Flow betweenness} $FB(i)$ -- is defined as the degree to which the maximum flow between all unordered pairs of vertices depends on an intermediary vertex $i$. Thus, the \textit{flow betweenness} $FB(i)$ is defined as~\citep{Freeman:1991un}:
\begin{equation}
\label{chp6_flow_betweenness}
FB(i)=C^W_{FB}(i)=\sum^n\limits_{\substack{{j=1}\\ j\neq i}}\sum^n\limits_{\substack{{k=1}\\ k\neq {i,j}}} f_{jk}(i),
~\end{equation}
where $f_{jk}(i)$ is the maximum flow from $j$ to $k$ that passes through vertex $i$.\\
%

\textit{Flow closeness} $FC(i)$ --  
although \citet{Freeman:1991un} did not formally define \textit{flow closeness} as a centrality measure, we can \textit{partially} extend this as the maximum flow between one vertex $i$ and the rest of the network, as:

\begin{equation}
\label{chp6_flow_closeness}
FC(i)=C^W_{FC}(i)=\sum^n_{j=1} f_{ij},
\end{equation}
where $f_{ij}$ represents the maximum flow from vertex $i$ to vertex $j$. It is important to highlight that this measure has a poor ability to distinguish which are the vertex(ices) closest to every other vertices when in presence of special situations, such as when the network has a star topology (one central vertex -- hub -- and the remaining $n-1$ vertices connected to it) in which all links have the same strength~\citep{Gomez:2013ee}.

\section{Centrality Measures for Interval-Weighted Networks}
\label{chp6_Sec_centrality_measures_IWN}

In classical network flow theory, a capacity $c(i,j)$ is associated with each edge between vertex vertex $i$ and $j$, denoting the maximum amount that can flow on the edge and a lower bound $l(i,j)$ representing the minimum amount that must flow on the edge~\citep{Ahuja:1993uh}. Nevertheless, the maximum flow problem is restricted by flow bounds considering only the maximum flow capacity $c(i,j)$ of an edge between each pair of vertices $i$ and $j$, thus assuming that this capacity is constant. However, in real-world applications, these capacities may vary within ranges rather than being constants~\citep{Ahuja:1993uh}. To better model such variability on an edge, instead of using constants, we represent flow capacities as intervals~\citep{Hu:2008vt,Sengupta:2009wk,Hossain:2010ur,Hossain:2010vq,Bozhenyuk:2017vd}. An interval representation of these capacities allows taking into account the variability observed in the original network, thereby minimizing the loss of information.
In what follows, we extend the degree, flow betweenness and flow closeness centrality measures to the general case of interval-weighted networks. First, we introduce the \textit{Interval-Weighted Degree} (IWD), extending~\citet{Opsahl:2010in} concept of a tuning parameter to give relevance to both edge weights and number of edges attached to a vertex. Secondly, based on capacity flow networks, using FF \textit{max-flow} algorithm~\citep{Ford:26m8xm4j}, we present the \textit{Interval-Weighted Flow Closeness} (IWFC) and \textit{Interval-Weighted Flow Betweenness} (IWFB).

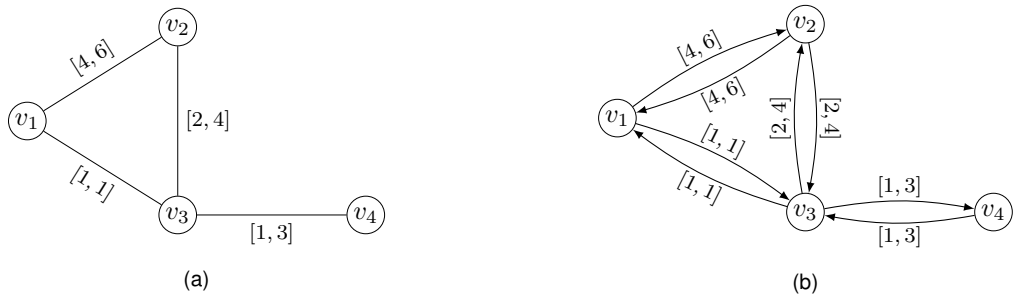
\begin{figure}[H]
			\centering
			\begin{subfigure}[c]{0.5\linewidth}
			\centering
			\begin{tikzpicture}[inner sep=0pt, minimum size=5mm, auto,
   				node_style/.style={draw,circle,line width=.1mm, font=\fontsize{10}{10}\selectfont},
   				edge_style/.style={draw=black, ultra thick, line width=.1mm}]
    			\node[node_style] (v1) at (0,1.25) {$v_1$};
    			\node[node_style] (v2) at (2,2.5) {$v_2$};
    			\node[node_style] (v3) at (2,0) {$v_3$};
    			\node[node_style] (v4) at (4.5,0) {$v_4$};
    \draw[edge_style]  (v1) edge node[above,sloped,pos=0.5,font=\fontsize{8}{10}\selectfont] {$[4,6]$} (v2);
    \draw[edge_style]  (v1) edge node[below,sloped,pos=0.5,font=\fontsize{8}{10}\selectfont] {$[1,1]$} (v3);
    \draw[edge_style]  (v2) edge node[right=0.1,pos=0.5,font=\fontsize{8}{10}\selectfont] {$[2,4]$} (v3);
    \draw[edge_style]  (v3) edge node[below,pos=0.5,font=\fontsize{8}{10}\selectfont] {$[1,3]$} (v4);
			\end{tikzpicture}
			\caption{}
			\end{subfigure}%
			\centering
			\begin{subfigure}[c]{0.5\linewidth}
			\centering
			\begin{tikzpicture}[inner sep=0pt, minimum size=5mm, auto,
   				node_style/.style={draw,circle,line width=.1mm, font=\fontsize{10}{10}\selectfont},
   				edge_style/.style={draw=black, ultra thick, line width=.1mm}]
    			\node[node_style] (v1) at (0,1.25) {$v_1$};
    			\node[node_style] (v2) at (2.5,2.5) {$v_2$};
    			\node[node_style] (v3) at (2.5,0) {$v_3$};
    			\node[node_style] (v4) at (5,0) {$v_4$};
				\draw[-latex,bend left=10]  (v1) edge node[above=-2pt,sloped,pos=0.5,font=\fontsize{8}{10}\selectfont] {$[4,6]$} (v2);    
    			\draw[-latex,bend left=10]  (v2) edge node[below=-2pt,sloped,pos=0.5,font=\fontsize{8}{10}\selectfont] {$[4,6]$} (v1);
				\draw[-latex,bend left=10]  (v1) edge node[above=-2pt,sloped,pos=0.5,font=\fontsize{8}{10}\selectfont] {$[1,1]$} (v3);    
    			\draw[-latex,bend left=10]  (v3) edge node[below=-2pt,sloped,pos=0.5,font=\fontsize{8}{10}\selectfont] {$[1,1]$} (v1);
    			\draw[-latex,bend left=10]  (v2) edge node[above=-2pt,sloped,pos=0.5,font=\fontsize{8}{10}\selectfont] {$[2,4]$} (v3);    
    			\draw[-latex,bend left=10]  (v3) edge node[above=-2pt,sloped,pos=0.5,font=\fontsize{8}{10}\selectfont] {$[2,4]$} (v2);
				\draw[-latex,bend left=10]  (v3) edge node[above=-2pt,sloped,pos=0.5,font=\fontsize{8}{10}\selectfont] {$[1,3]$} (v4);    
    			\draw[-latex,bend left=10]  (v4) edge node[below=-2pt,sloped,pos=0.5,font=\fontsize{8}{10}\selectfont] {$[1,3]$} (v3);
			\end{tikzpicture}
			\caption{}
			\end{subfigure}%
\caption[Example of the transformation of an undirected interval--weighted network into a directed interval--weighted network]{(a) Undirected Interval--Weighted Network, and its corresponding (b) Directed Interval--Weighted Network~\citep{Schroeder:2004uv}.}
\label{chp6_fig:conversion_to_direct_net}			
\end{figure}

\paragraph{Conversion of an interval-weighted undirected network into its corresponding direct version}
Before generalizing the centrality measures discussed in Section~\ref{chp6_Sec_centrality_weighted_networks} to IWN, as mentioned in Note~\ref{note_network_conversion_undirected-directed}, it is first necessary to transform an undirected interval-weighted network into a direct one. Figure~\ref{chp6_fig:conversion_to_direct_net} shows an undirected interval-weighted network and its transformation into a directed interval-weighted network (for the sake of simplicity, in future representations of IWN, only one undirected interval-weighted edge will be represented, as shown in Figure~\ref{chp6_fig:conversion_to_direct_net}.a).

\paragraph{Flows in undirected interval-weighted networks}
The generalization to interval-weighted networks (IWN) of Freeman's betweenness and closeness centrality measures~\citep{Freeman:1991un}, according to the methodology based on ``flow networks'' discussed in Section~\ref{chp6_Sec_centrality_weighted_networks}, faced two major drawbacks proper of interval arithmetic~\citep{Moore:2009uc}:
\begin{itemize}
\item firstly, because of the  \textit{non-existence of inverse elements}. The generalization of \textit{closeness} and \textit{betweenness} to weighted networks done by \citet{Newman:2001kc} and \citet{Brandes:2001wm}, respectively, inverted the edge weights to consider them as \textit{costs} and then applied the Dijkstra's~(\citeyear{Dijkstra:1959vb}) shortest path algorithm (the least costly path connecting to vertices was the shortest path between them). Thus, the identification and length of the shortest paths is based on the sum of the inverted edge weights and is defined as: $d^W(i,j)=\min\left(\frac{1}{w_{ih}}+\dots+\frac{1}{w_{h'j}}\right)$, where $h$ and $h'$ are intermediary vertices on paths between vertices $i$ and $j$;
\item secondly, to what is known as \textit{interval dependency problem}. The use of flow capacities as intervals raises some difficulties when calculating the shift of a flow on an augmented path, since the real number arithmetic property of additive inverse (i.e., if $b=c-a$, then $a+b=c$) is not valid for interval subtraction, e.g., $[1,5]+[1,3]=[2,8]$, but $[2,8]-[1,3]=[-1,7]\neq [1,5]$ (see section~\ref{Interval_Analysis} for a detailed explanation about interval arithmetic).
\end{itemize}

\paragraph{Lexicographic order}

To circumvent the above mentioned difficulty, the \textit{lexicographic order} was used to prove that the \textit{maximum flow} is obtained with the maximum flow values at each edge, and the minimum flow with the minimum flow values at each edge (see Appendix A for details).

\subsection{Interval-Weighted Degree (IWD)}
\label{chp6_SubSec_degree_IWN}

The extension of the \textit{degree} to the case of an \textit{interval-weighted network} is done first by considering only the vertices strength, i.e., the sum of the edge interval-weights \citep{2004PNAS..101.3747B}, and secondly by taking into consideration both the number of edges and the edges strength by introducing a tuning parameter $\alpha$ \citep{Opsahl:2010in}. The following definitions express these concepts:

\begin{definition}
Using (\ref{chp6_weighted-degree_1}) and (\ref{chp6_weighted-degree_1_opsahl}), the generalization of degree to an IWN is formalized as:
\begin{equation}
\label{chp6_IW_degree}
IWD(i)=s^{IW}_i=C^{IW}_D(i)=\sum^n_{j=1} \left[\underline{w}_{ij},\overline{w}_{ij}\right],
\end{equation}
where $\left[\underline{w}_{ij},\overline{w}_{ij}\right]$ are the interval--weights $(\overline{w}_{ij}\geqslant \underline{w}_{ij}>0)$.
\end{definition}

\begin{definition}
The generalization of~\citet{Opsahl:2010in} approach, which consists in the inclusion of a tuning parameter, $\alpha$, to address the relative importance of the number of edges compared to edge interval-weights, is formalized as:
\begin{equation}
\label{chp6_IW_degree_opsahl}
IWD^{\alpha}(i)=C^{IW\alpha}_D (i)=k^{(1-\alpha)}_i \left(\sum^n_{j=1} \left[\underline{w}_{ij},\overline{w}_{ij}\right]\right)^{\alpha}=k^{(1-\alpha)}_i \left(s^{IW}_i\right)^{\alpha}.
\end{equation}
where $k_i$ is the number of vertices that a focal vertex $i$ is connected to.
\end{definition}

The following is an example of our approach to degree in an IWN.

\begin{example}[Interval-Weighted Degree -- IWD]\label{chp6_Example_Degree}
Given an interval-weighted network (Figure~\ref{chp6_Example_IWD_Degree_network}), the IWD for the different $\alpha$ benchmark values is shown in Table~\ref{chp6_Example_IWD_Degree_table}.

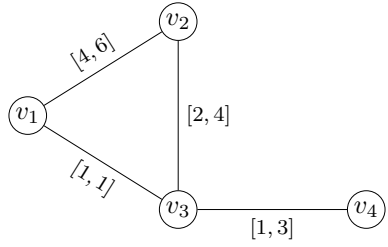
\begin{figure}[H]
	\centering
	\begin{tikzpicture}[inner sep=0pt, minimum size=5mm, auto,
		node_style/.style={draw,circle,line width=.1mm, font=\fontsize{10}{10}\selectfont},
		edge_style/.style={draw=black, ultra thick, line width=.1mm}]
		\node[node_style] (v1) at (0,1.25) {$v_1$};
		\node[node_style] (v2) at (2,2.5) {$v_2$};
		\node[node_style] (v3) at (2,0) {$v_3$};
		\node[node_style] (v4) at (4.5,0) {$v_4$};
	    \draw[edge_style]  (v1) edge node[above,sloped,pos=0.5,font=\fontsize{8}{10}\selectfont] {$[4,6]$} (v2);    							\draw[edge_style]  (v1) edge node[below,sloped,pos=0.5,font=\fontsize{8}{10}\selectfont] {$[1,1]$} (v3);
	    \draw[edge_style]  (v2) edge node[right=0.1,pos=0.5,font=\fontsize{8}{10}\selectfont] {$[2,4]$} (v3);
	    \draw[edge_style]  (v3) edge node[below,pos=0.5,font=\fontsize{8}{10}\selectfont] {$[1,3]$} (v4);
	\end{tikzpicture}
	\caption{Interval--Weighted Network}
	\label{chp6_Example_IWD_Degree_network}
\end{figure}

Analysing Table~\ref{chp6_Example_IWD_Degree_table}, we conclude that when the degree outcome is solely based on the number of edges $(\alpha=0)$, the edge weights are ignored (i.e., the degree is the same as if the network were binary), then vertex $v_3$ is the most central. The same occurs when we consider a tuning parameter between $0$ and $1$ (e.g. $\alpha=0.5$), which means that the degree would positively value both the number of edges and the edge weights, with weights varying in $[3.46,4.90]$.

However, when the degree is based only on edge weights $(\alpha=1)$, vertex $v_2$ becomes the most central one with the degree varying in $[6,10]$. The same outcome is obtained if the tuning parameter is above one (e.g. $\alpha=1.5$), which positively values edge strength and negatively values the number of edges $[10.39,22.36]$.\\

\begin{table}[H]
\caption{Interval--Weighted Degree values for the IWN in Figure~\ref{chp6_Example_IWD_Degree_network}, for the benchmark values $\alpha=0,0.5,1,1.5$.}
\label{chp6_Example_IWD_Degree_table}
\centering
\renewcommand{\arraystretch}{1}
\setlength{\tabcolsep}{2.1pt}
\fontsize{8}{10}\selectfont
\begin{tabular}{M{1cm}?M{1.5cm}|M{1cm}?M{1.5cm}|M{1cm}?M{1.5cm}|M{1cm}?M{1.9cm}|M{1cm}}
\thickhline
				 & \multicolumn{8}{c}{Tuning parameter $(\alpha)$} \\\cline{2-9} 
                        & \multicolumn{2}{c?}{\cellcolor{gray!20}$\alpha=0$}        & \multicolumn{2}{c?}{\cellcolor{gray!20}$\alpha=0.5$}       & \multicolumn{2}{c?}{\cellcolor{gray!20}$\alpha=1$}        & \multicolumn{2}{c}{\cellcolor{gray!20}$\alpha=1.5$}         \\\cline{2-9}
\multirow{2}{*}{Vertex}   & \makecell{Interval\\degree} & \makecell{Interval\\rank} & \makecell{Interval\\degree} & \makecell{Interval\\rank} & \makecell{Interval\\degree} & \makecell{Interval\\rank} & \makecell{Interval\\degree} & \makecell{Interval\\rank} \\\thickhline  
$v_1$                      & $[2,2]$   & $2\textsuperscript{nd}$   & $[3.16, 3.74]$ & $3\textsuperscript{rd}$    & $[5, 7]$  & $3\textsuperscript{rd}$  & $[7.91, 13.10]$  & $2\textsuperscript{nd}$  \\\hline
$v_2$                      & $[2,2]$   & $2\textsuperscript{nd}$   & $[3.46, 4.47]$ & $2\textsuperscript{nd}$    & $\mathbf{[6, 10]}$ & $\mathbf{1\textsuperscript{st}}$  & $\mathbf{[10.39, 22.36]}$ & $\mathbf{1\textsuperscript{st}}$  \\\hline
$v_3$                      & $\mathbf{[3,3]}$	 & $\mathbf{1\textsuperscript{st}}$   & $\mathbf{[3.46, 4.90]}$ & $\mathbf{1\textsuperscript{st}}$    & $[4, 8]$  & $2\textsuperscript{nd}$  & $[4.62, 13.06]$  & $3\textsuperscript{rd}$   \\\hline
$v_4$                      & $[1,1]$   & $4\textsuperscript{th}$   & $[1.00, 1.73]$ & $4\textsuperscript{th}$    & $[1, 3]$  & $4\textsuperscript{th}$  & $[1.00, 5.20]$   & $4\textsuperscript{th}$     \\   \thickhline      
\end{tabular}
\end{table}
\end{example}


\subsection{Interval-Weighted Flow Centrality Measures}
\label{chp6_SubSec_Flow_measures_IWN}

%
\begin{definition}[Interval-Weighted Flow Betweenness (IWFB)]
Using (\ref{chp6_flow_betweenness}), the generalization of flow betweenness (FB) to an IWN is formalized as:
\begin{equation}
\label{chp6_IW_flow_betweenness}
IWFB(i)=C^{IW}_{FB}(i)=\left[\sum^n\limits_{\substack{{j=1}\\ j\neq i}}\sum^n\limits_{\substack{{k=1}\\ k\neq {i,j}}} \underline{f}_{jk}(i), \sum^n\limits_{\substack{{j=1}\\ j\neq i}}\sum^n\limits_{\substack{{k=1}\\ k\neq {i,j}}} \overline{f}_{jk}(i)\right],
\end{equation}
where $\underline{f}_{jk}(i)$ and $\overline{f}_{jk}(i)$ are the minimum and the maximum flow, for the lower and upper bounds of the weighted intervals, from $j$ to $k$ that pass through vertex $i$, respectively.\\
\end{definition}

%
\begin{definition}[Interval-Weighted Flow Closeness (IWFC)]
Using (\ref{chp6_flow_closeness}), the generalization of flow closeness (FC) to an IWN is formalized as:
\begin{equation}
\label{chp6_IW_flow_closeness}
IWFC(i)=C^{IW}_{FC}(i)=\left[\sum^n_{j=1} \underline{f}_{ij}, \sum^n_{j=1} \overline{f}_{ij}\right],
\end{equation}
where $\underline{f}_{ij}$ and $\overline{f}_{ij}$ are the minimum and the maximum flow for the lower and upper bounds of the weighted intervals between vertices $i$ and $j$, respectively.
\end{definition}

Below is an example of the values obtained for the two centrality measures on an interval-weighted flow network.\\

\begin{example}[Interval-Weighted Flow Centrality Measures, Betweenness (IWFB) and Closeness (IWFC)]\label{chp6_Example_IWFB-IWFC}
Given the interval-weighted network used in Example~\ref{chp6_Example_Degree}, Figure~\ref{chp6_Example_IWD_Degree_network}, the values of the IWFB and IWFC are shown in Table~\ref{chp6_Example_IWFB-IWFC_table}.

\begin{table}[H]
\centering
\caption{Interval-Weighted Flow Centrality measures for the IWN in Figure~\ref{chp6_Example_IWD_Degree_network}.}
\label{chp6_Example_IWFB-IWFC_table}
\renewcommand{\arraystretch}{1}
\setlength{\tabcolsep}{2.1pt}
\fontsize{8}{10}\selectfont
\begin{tabular}{M{1cm}?M{1cm}|M{1cm}|M{1cm}|M{1cm}?M{1.25cm}|M{1.25cm}?M{1.25cm}?M{1.5cm}?M{1.25cm}|M{1.5cm}}
\thickhline
					& \multicolumn{4}{c?}{Ford \& Fulkerson \textit{max-flow}\textsuperscript{a}}  & \multicolumn{2}{c?}{\textit{max-flow} (all pairs)\textsuperscript{b}} & \multicolumn{2}{c?}{\textbf{Flow Betweenness}} & \multicolumn{2}{c}{\textbf{Flow Closeness}} \\ \hhline{~----------}
Vertex          & $v_1$    & $v_2$   & $v_3$    & $v_4$   & \textit{max-flow} & \textit{max-flow} rank & \cellcolor{gray!20} \textbf{IWFB\textsuperscript{c}}    & IWFB rank      & \cellcolor{gray!20} \textbf{IWFC\textsuperscript{d}}        & IWFB rank  \\ \thickhline 
$v_1$   & $[0,0]$ & $[5,7]$ & $[3,5]$ & $[1,3]$ & $[5, 11]$ & $3\textsuperscript{rd}$   & \cellcolor{gray!20}$[1, 1]$  & $3\textsuperscript{rd}$  & \cellcolor{gray!20} $\mathbf{[9, 15]}$  & $1\textsuperscript{st}$      \\ \hline
$v_2$  & $[5,7]$ & $[0,0]$ & $[3,5]$ & $[1,3]$ & $[5, 11]$  & $3\textsuperscript{rd}$  & \cellcolor{gray!20}$[2, 6]$  & $2\textsuperscript{nd}$    & \cellcolor{gray!20} $\mathbf{[9, 15]}$ & $1\textsuperscript{st}$ \\ \hline
$v_3$  & $[3,5]$ & $[3,5]$ & $[0,0]$ & $[1,3]$ & $[7, 13]$  & $2\textsuperscript{nd}$ & \cellcolor{gray!20} $\mathbf{[3, 7]}$  & $1\textsuperscript{st}$   & \cellcolor{gray!20}$[7, 13]$  & $3\textsuperscript{rd}$    \\ \hline
$v_4$  & $[1,3]$ & $[1,3]$ & $[1,3]$ & $[0,0]$ & $[11, 17]$ & $1\textsuperscript{st}$  & \cellcolor{gray!20}$[0, 0]$  & $4\textsuperscript{th}$   & \cellcolor{gray!20}$[3, 9]$   & $4\textsuperscript{th}$  \\ \thickhline 
\end{tabular}
\begin{tablenotes}
      \tiny
      \item {\textsuperscript{a} Ford \& Fulkerson's \textit{max-flow} between vertices.}
      \item {\textsuperscript{b} {max-flow} between all pairs of vertices, where the vertex $v_i$ is neither a source or a sink.}
      \item {\textsuperscript{c} Interval-Weighted Flow Betweenness centrality.}
      \item {\textsuperscript{d} Interval-Weighted Flow Closeness centrality.}
\end{tablenotes}
\end{table}

Regarding to \textit{betweenness}, Table~\ref{chp6_Example_IWFB-IWFC_table} shows that vertex $v_3$ has the higher betweenness centrality varying in $IWFB=[3,7]$. This means that, of the total maximum flow between all pairs of vertices, where the vertex $v_3$ is neither a source or a sink $[7,13]$, a flow between $3$ and $7$ must pass through vertex $v_3$.

On the contrary, relatively to \textit{closeness}, Table~\ref{chp6_Example_IWFB-IWFC_table} shows that vertices $v_1$ and $v_2$ have the highest closeness centrality, varying between $9$ and $15$.
\end{example}

Below in Figure~\ref{chp6_fig:max-flow calculations-v3} is depicted an example that illustrates how to obtain the \textit{max-flow} $[7,13]$ and the \textit{interval weighted flow betweenness} (IWFB) $[3,7]$ values for vertex $v_3$, when $v_3$ is neither a \textit{source} nor a \textit{sink}, as in Table~\ref{chp6_Example_IWFB-IWFC_table}.

\begin{figure}[H]
	\centering
    	\includegraphics[scale=0.72, clip, trim={0cm 9cm 0cm 8.9cm}]{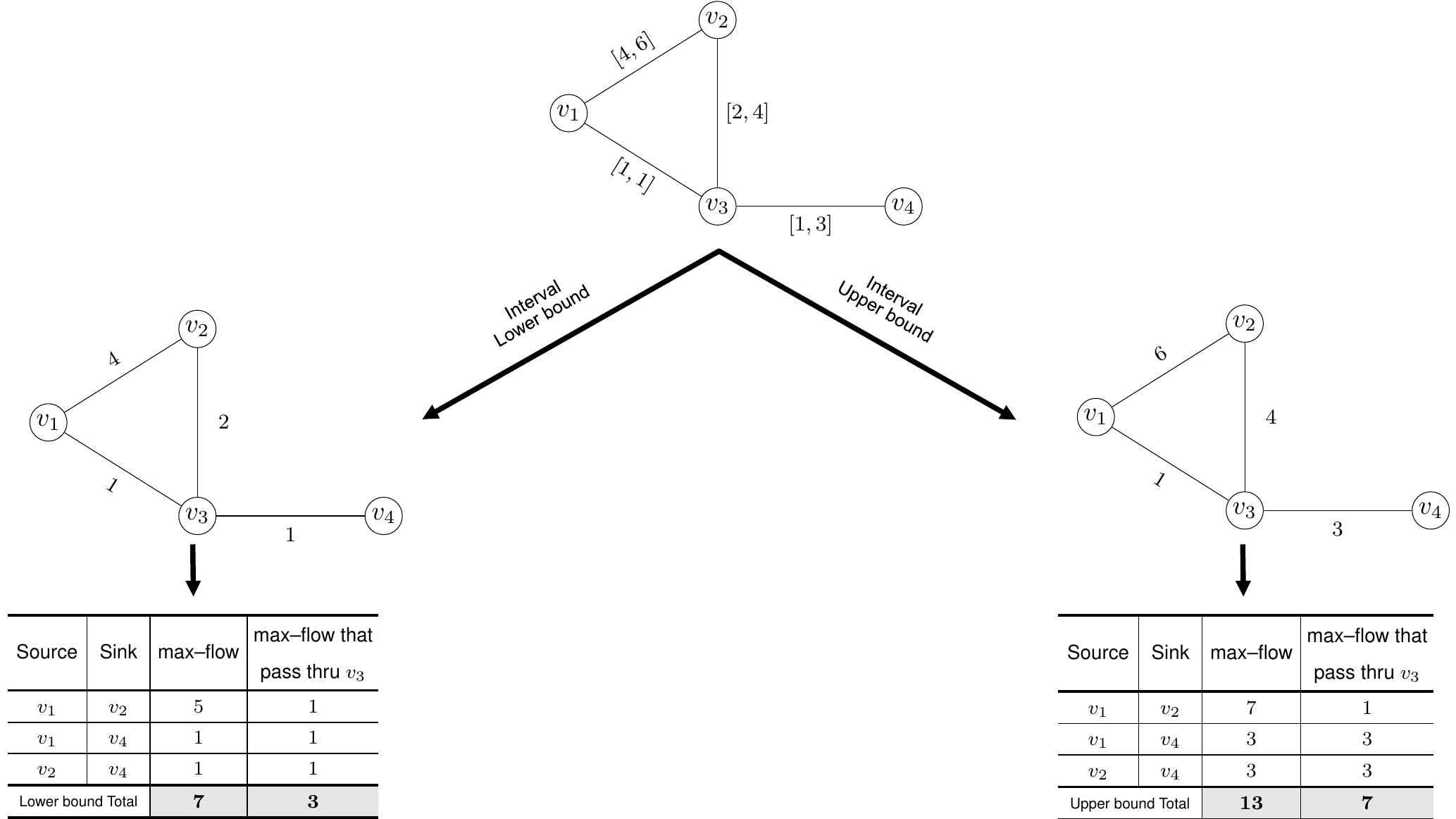}
	\caption{In the center is the interval-weighted network, on the left and right side are shown the weighted networks and the tables with the max-flow values for vertex $v_3$, when $v_3$ is neither a \textit{source} nor a \textit{sink} for the intervals \textit{lower} and \textit{upper} bounds, respectively.}
	\label{chp6_fig:max-flow calculations-v3}
\end{figure}

Next, applications to two real-world interval-weighted networks further illustrates the proposed approaches.

\section{Applications}
\label{applications}

In recent years, centrality measures have often been used in complex networks representing territorial units as tools to identify the central units~\citep{DeMontis:2007iu,DeMontis:2011db,Cheng:2015in}. We present the application of our centrality measures approach to two real-world interval-weighted networks. The first network represents the movements of daily commuters in mainland Portugal\footnote{\url{https://www.ine.pt/xportal/xmain?xpid=INE&xpgid=ine_base_dados&bdpagenumber=2&contexto=bd&bdtemas=1115&bdsubtemas=111514&bdfreetext=pendulares&xlang=pt}.} (by all means of transportation) between the 23 NUTS 3 Regions\footnote{NUTS--Nomenclature of Territorial Units for Statistics~\citep{Eurostat:2016a}.} (henceforth, the ``Interval-Weighted Commuters Network (IWCN)'') (source: INE -- Statistics Portugal, Census 2011). The second application focuses on annual Merchandise trade (detailed products, exports in thousands of US dollars) between 28 European countries from 2003 to 2015 (henceforth, the ``Interval-Weighted Trade Network (IWTN)''), analysing the commercial communities that emanate between these countries for the thirteen year period considered (henceforth, the ``Interval--Weighted Trade Network (IWTN)'')~\citep{UNCTAD:2016a}.

\subsection{Network of Portuguese commuters}
\label{applied_exemple_1}

\begin{figure}[ht]
	\centering
    	\includegraphics[scale=0.8, clip, trim={10cm 7.5cm 9.5cm 4cm}]{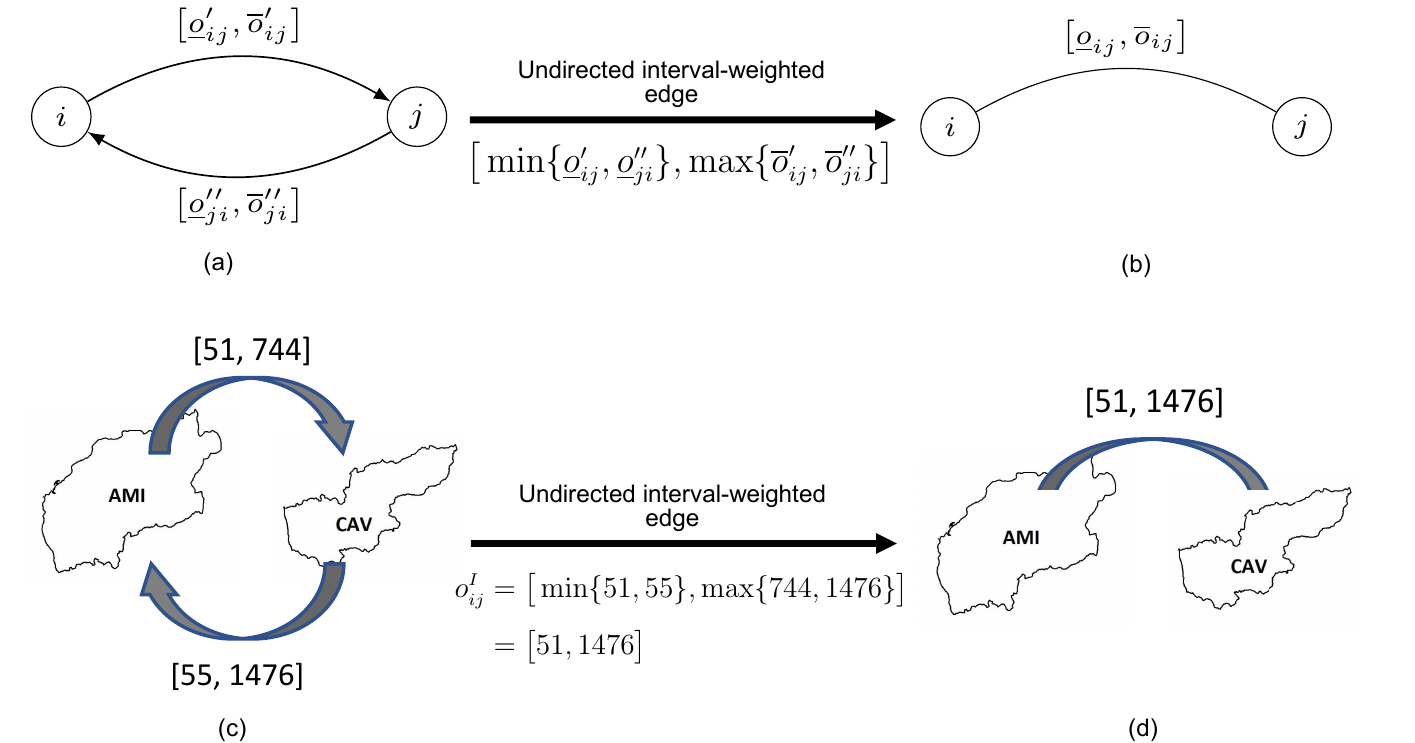}
	\caption{Conversion of directed interval-weighted edges into an undirected interval-weighted edge. (a) Bidirectional interval flows $i\to j$ and $j\to i$ between NUTS 3 $i$ and $j$, (b) Undirected interval flow between NUTS 3 $i$ and $j$, (c) and (d) are an example extracted from the real data.}
	\label{chp7_fig:directed_to_undirected_edges}
\end{figure}

According to various authors, the \textit{flows} of daily commuters can be conceived as a network \citep{DeMontis:2007iu,Patuelli:2007ks,DeMontis:2011db,DeMontis:2013ho,DeLeo:2013do,Cheng:2015in,2016JSMTE..03.3404X,Barbosa:2018ki,Zeng:2018jm,Spadon:2019bv}. Hence, each \textit{vertex} of the Interval-Weighted Commuters Network (IWCN) corresponds to a given NUTS 3 (which in turn represents the aggregation of commuter flows between the municipalities that constitute it) and the \textit{edges} represent intervals ranging between the \textit{minimum} flow \textit{larger than 50 commuters} and the \textit{maximum} flow of commuters between the corresponding NUTS 3. As represented in Figure~\ref{chp7_fig:directed_to_undirected_edges}a, the interval of commuters flow from NUTS $i\to j$ may be  different from the one of $j\to i$. Therefore, the elements $o^I_{ij}$ of the symmetric interval-weighted adjacency matrix, $O^I$, denote the maximum variability of the \textit{bi-directional} flows $ij$ and $ji$ between the NUTS $i$ and $j$ (Figure\ref{chp7_fig:directed_to_undirected_edges}b): $o^I_{ij}=\big[\min\{\underline{o}'_{ij},\underline{o}''_{ji}\},\max\{\overline{o}'_{ij},\overline{o}''_{ji}\}\big]=\big[\underline{o}_{ij},\overline{o}_{ij}\big]$. The option for this representation of flows is related to the fact that we do not want to study the orientation of these daily commuter fluxes, but just quantify the reciprocal attractiveness of the NUTS 3 pairs~\citep{DeMontis:2013ho}. This kind of aggregation when the data are recorded at the same point in time and the statistical units to be analysed are not those for which the data was originally recorded, but constitute specific groups of those (level higher than the one at which the data was originally collected), is called \textit{contemporary aggregation}~\citep{Brito:2014es}.

The adjacency matrix elements are null, $o^I_{ij}=[0,0]$, when there is no commuter flow greater than 50 daily movements between NUTS 3 $i$ and $j$. By definition, we assume that there are no commuter flows within each NUTS 3, i.e., the network has no loops at initial  vertices, which implies that the diagonal of the interval-weighted adjacency matrix consists of degenerate intervals with the value zero, $o^I_{ii}=[0,0]$.

\begin{figure}[H]
    \centering
    \begin{subfigure}[t]{0.48\linewidth}
        \centering
        \includegraphics[width=0.5\linewidth,clip, trim={0cm 0cm 0cm 0cm}]{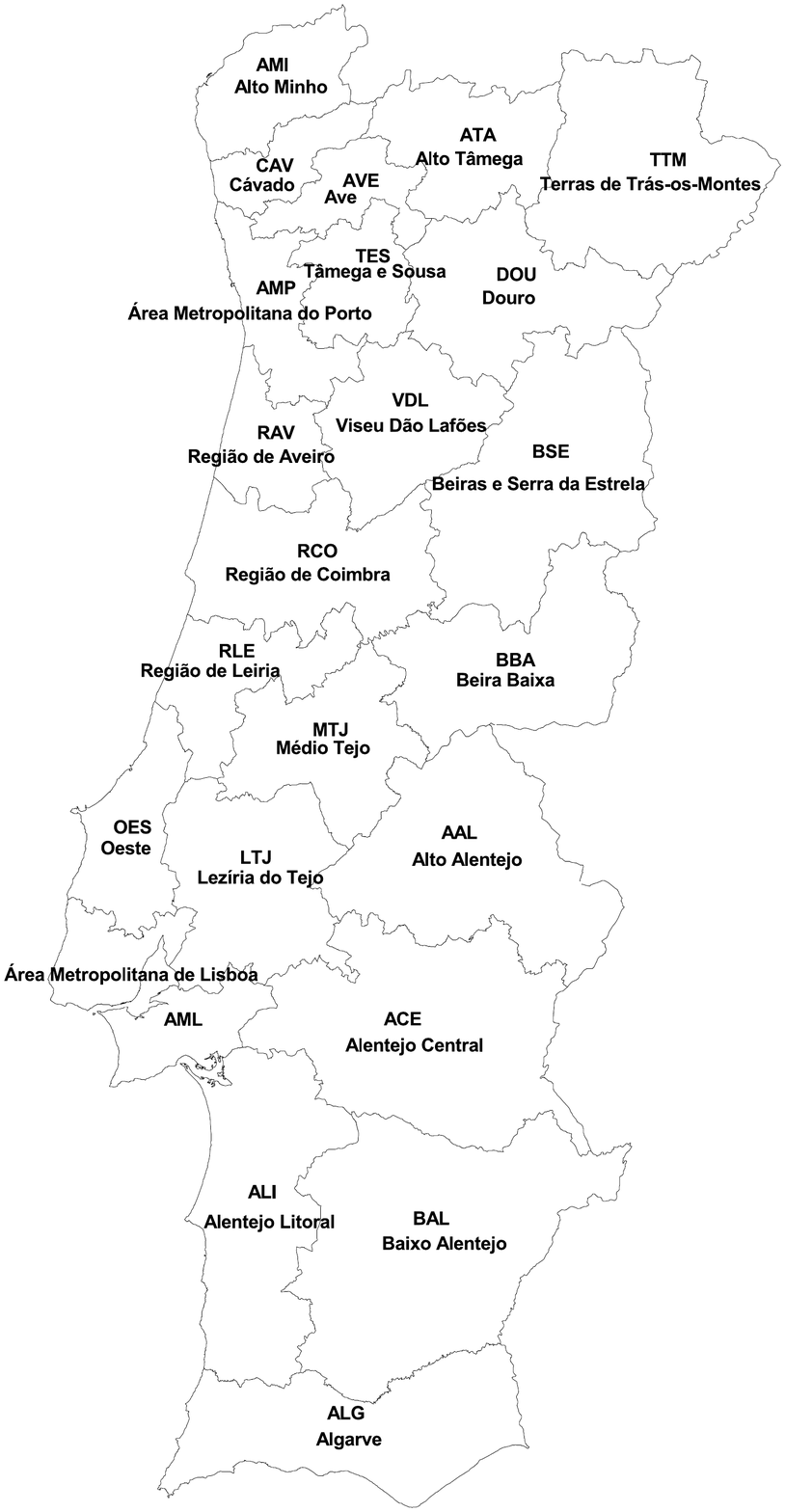}
        \caption{}
        \label{chp7_fig:Map_PT}
    \end{subfigure}
    \begin{subfigure}[t]{0.45\linewidth}
        \centering
        \includegraphics[width=0.5\linewidth, clip, trim={0cm 0cm 0cm 0cm}]{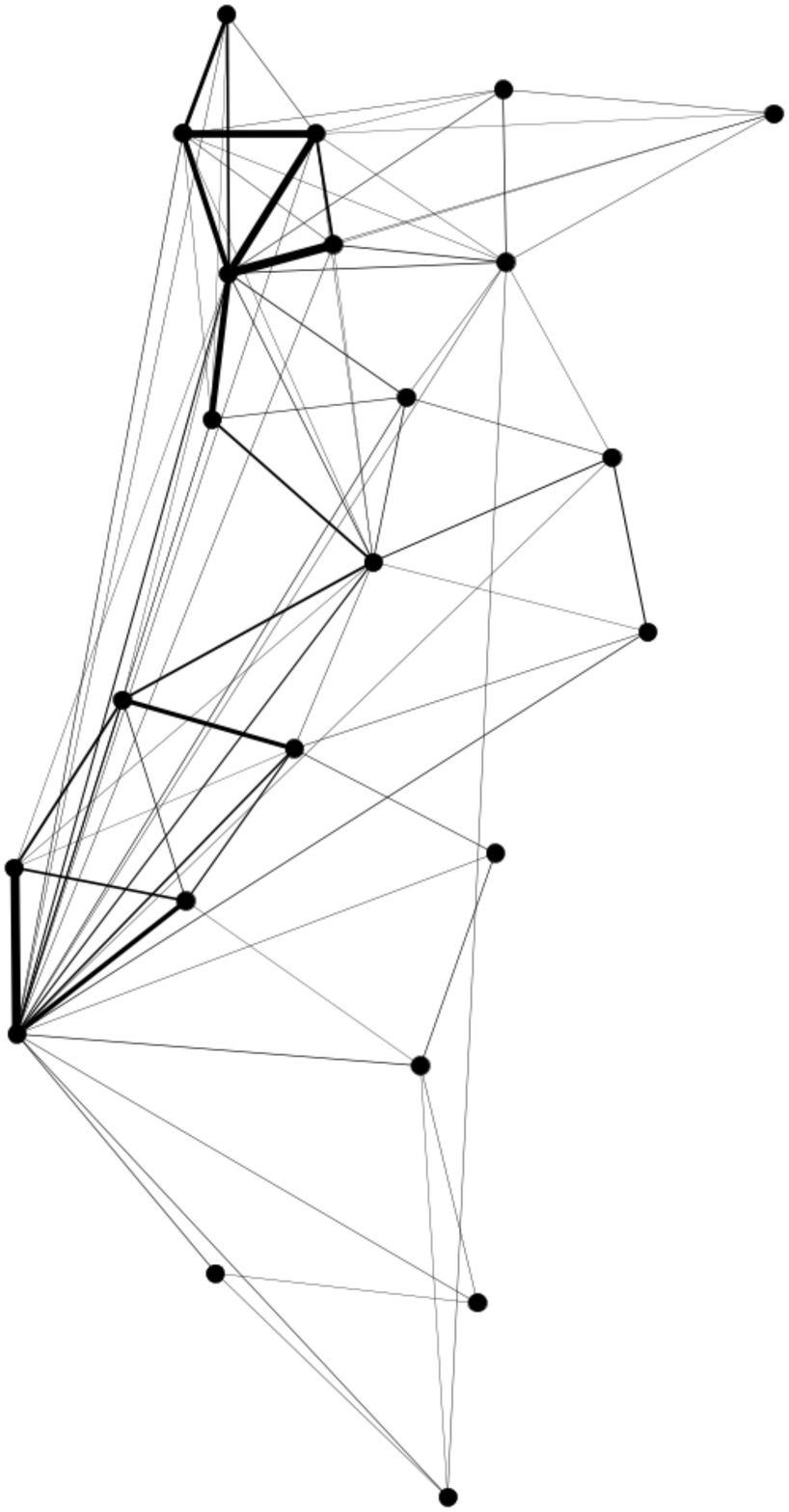}
        \caption{}
        \label{chp7_fig:Map_PTNet}
    \end{subfigure}
    \caption[Geographic representation of Portuguese NUTS 3 and the correspondent weighted network]{(a) Geographic representation of Portuguese NUTS 3, and (b) Topologic representation of the Portuguese NUTS 3 interval-weighted commuters network (IWCN).}
    \label{chp7_fig:Map_PTandNet}
\end{figure}

Figure~\ref{chp7_fig:Map_PTandNet} shows the geographical distribution of NUTS 3 in mainland Portugal (Figure~\ref{chp7_fig:Map_PT}), and the corresponding network of commuting movements between these NUTS 3, weighted by intervals denoting the maximum variability (Figure\ref{chp7_fig:Map_PTNet})\footnote{For the sake of visualization, we chose not to represent the intervals on the network edges, such as it is depicted in Figure~\ref{chp7_fig:directed_to_undirected_edges}d.}. This network has 23 vertices and 80 edges and is therefore considered a small network with low density (considering the intervals midpoints: $\text{graph density}=0.316$, $\text{diameter}=3$, $\text{average degree}=6.96$). For ease of reading, hereinafter we will only refer to Portugal instead of ``mainland Portugal''.

\subsubsection{Results -- interval-weighted commuters network (IWCN)}
\label{results_IWCN}


In this section, applying the new centrality measures that incorporate the \textit{tuning parameter} $\alpha$ of Opsahl's et.al.~(\citeyear{Opsahl:2010in}) for \textit{degree} and the \textit{flow capacities} of Freeman et.al.~(\citeyear{Freeman:1991un}) for \textit{betweenness} and \textit{closeness}, we aim at identifying which are the critical (most central) vertices in the interval-weighted network described above, i.e., which are the most central NUTS 3 in the country.



\paragraph{Interval-Weighted Degree Centrality (IWD)}
Table~\ref{chp7_Results_IWD_Degree_table_CommutersNetwork} ranks in descending order the 23 NUTS 3 according to the \textit{degree centrality} interval score for different values of the tuning parameter $\alpha$. Highlighted in gray are the cases where there was a shift in the interval rank classification with the change of the $\alpha$ value. To simplify our analysis, we will focus only on the two main regions, AML (Lisbon Metropolitan Area) and AMP (Porto Metropolitan Area). As expected, AML and AMP are the most central NUTS 3, irrespective of $\alpha$. A closer look however reveals that when the degree outcome is solely based on the number of edges, $\alpha=0$ (the edge weights are ignored and the degree is the same as if the network were binary), AML is the most central NUTS 3 and AMP comes in second place. The same occurs when we consider a tuning parameter such that $0<\alpha<1$ $(\alpha=0.5)$ (the degree would positively value both the number of edges and the edge weights), with weights varying approximately in $[159,436]$.\\
However, when the degree is based only on edge weights $(\alpha=1)$, AMP becomes the most central NUTS 3 with the degree varying in $[939,12997]$\footnote{As defined in Section~\ref{chp6_Sec_centrality_weighted_networks}, (\ref{chp6_weighted-degree_1_opsahl}), this measure is the product of the number of vertices that a focal vertex is connected to, by the average weight of these vertices adjusted by a tuning parameter $\alpha$. Thus, for this particular interval, we conclude that the \textit{average weight} of the NUTS 3 attached to AMP varies within $\left[\frac{939}{20},\frac{12997}{20}\right]=[46.95,649.85]$.}. Thus, we may conclude that AML is connected with more NUTS 3, but AMP despite having fewer connections, tends to have connections that involve more commuters.
The same outcome is obtained if the tuning parameter is above one $(\alpha=1.5)$, which positively values edge strength and negatively values the number of edges $[7980,410954]$.

For the remaining NUTS 3 marked in gray there are also shifts in the degree interval rank classification with the change of the $\alpha$ parameter, which means that this measure, in fact, considers both the number of edges and the edges strength as well as being sensitive to the \textit{average edge weight of a vertex}. In particular, Oeste (OES) clearly climbs in the ranking for $\alpha\geqslant 1$, showing that although not connected to many regions, its connections involve a large number of commuters; the opposite is observed for the Douro region (DOU).

\begin{table}[t]
\centering
\begin{threeparttable}
\caption{Degree centrality for the Interval-Weighted Commuters network (NUTS 3 ranked in descending order of \textit{interval rank} for $\alpha=1$).}
\label{chp7_Results_IWD_Degree_table_CommutersNetwork}
\centering
\renewcommand{\arraystretch}{.85}
\setlength{\tabcolsep}{2.1pt}
\fontsize{7.5}{10}\selectfont
\begin{tabular}{M{1.25cm}|M{1.5cm}|M{1cm}|M{2cm}|M{1cm}|M{1.75cm}|M{1cm}|M{3cm}|M{1cm}}
\thickhline
				 & \multicolumn{8}{c}{Tuning parameter $(\alpha)$} \\\cline{2-9} 
                        & \multicolumn{2}{c|}{$\alpha=0$}        & \multicolumn{2}{c|}{$\alpha=0.5$}       & \multicolumn{2}{c|}{$\alpha=1$}        & \multicolumn{2}{c}{$\alpha=1.5$}         \\\cline{2-9}
\multirow{2}{*}{NUTS 3\tnote{a}}   & \makecell{Degree\\interval} & \makecell{Interval\\rank} & \makecell{Degree\\interval} & \makecell{Interval\\rank} & \makecell{Degree\\interval} & \makecell{Interval\\rank} & \makecell{Degree\\interval} & \makecell{Interval\\rank} \\\thickhline  
\rowcolor{gray!25}AMP	& $[13 , 13]$ &	 $2$	& $[110.49 , 411.05]$ &	$2$	& $[939  , 12997]$ &$1$	& $[7980.44 , 410953.80]$ &	$1$ \\\hline
\rowcolor{gray!25}AML	& $[20 , 20]$ &	  $1$	& $[159.19 , 436.30]$ & $1$	& $[1267 , 9518]$ &	$2$	& $[10084.40 , 207636.43]$ &	$2$ \\\hline
AVE	& $[9  , 9]$ &	$5$	& $[69.46   , 253.83]$ &	$3$	& $[536  , 7159]$ &	$3$	& $[4136.43  , 201909.75]$ &	$3$ \\\hline
CAV	& $[9  , 9]$ &	$5$	& $[69.91   , 249.27]$ &	$4$	& $[543  , 6904]$ &	$4$	& $[4217.73  , 191218.51]$ &	$4$ \\\hline
\rowcolor{gray!25}OES	& $[6  , 6]$ &	$11$	& $[46.15  , 178.43]$ &	$8$	& $[355  , 5306]$ &	$5$	& $[2730.66  , 157788.46]$ &	$5$ \\\hline
TES	& $[7  , 7]$ &	$7$	& $[51.91   , 189.04]$ &	$6$	& $[385  , 5105]$ &	$6$	& $[2855.24  , 137862.01]$ &	$6$ \\\hline
RLE	& $[7  , 7]$ &	$7$	& $[51.98   , 176.85]$ &	$7$	& $[386  , 4468]$ &	$7$	& $[2866.37  , 112880.97]$ &	$7$ \\\hline
\rowcolor{gray!25}RCO	& $[12 , 12]$ &	$3$	& $[92.04  , 210.09]$ &	$5$	& $[706  , 3678]$ &	$8$	& $[5415.22  , 64391.27]$ &	$11$ \\\hline
RAV	& $[7  , 7]$ &	$7$	& $[52.18   , 162.49]$ &	$9$	& $[389  , 3772]$ &	$9$	& $[2899.85  , 87560.56]$ &	$9$  \\\hline
MTJ	& $[7  , 7]$ &	$7$	& $[54.03   , 154.11]$ &	$10$	& $[417 , 3393]$ &	$10$	& $[3218.51 , 74701.06]$ &	$10$ \\\hline
\rowcolor{gray!25}LTJ	& $[5  , 5]$ &	$13$	& $[36.54  , 132.00]$ &	$12$	& $[267 , 3485]$ &	$11$	& $[1951.11 , 92006.64]$ &	$8$  \\\hline
\rowcolor{gray!25}AMI	& $[5  , 5]$ &	$13$	& $[37.22  , 115.20]$ & $13$	& $[277 , 2654]$ &	$12$	& $[2061.74 , 61145.76]$ &	$12$ \\\hline
\rowcolor{gray!25}DOU	& $[10 , 10]$ &	$4$	& $[76.22  , 124.74]$ &	$11$	& $[581 , 1556]$ &	$13$	& $[4428.58 , 19409.50]$ &	$13$ \\\hline
VDL	& $[6  , 6]$ &	$11$	& $[52.31  , 87.02]$ &	$14$	& $[456  , 1262]$ &	$14$	& $[3975.32 , 18302.63]$ &	$14$ \\\hline
BSE	& $[5  , 5]$ &	$13$	& $[36.26  , 79.56]$ &	$15$	& $[263  , 1266]$ &	$15$	& $[1907.43 , 20144.92]$ &	$15$ \\\hline
\rowcolor{gray!25}BBA	& $[4  , 4]$ &	$19$	& $[36.72  , 59.06]$ &	$18$	& $[337  , 872]$ &	$16$	& $[3093.25  , 12874.93]$ &	$16$ \\\hline
\rowcolor{gray!25}ATA	& $[5  , 5]$ &	$13$	& $[44.05  , 58.40]$ &	$16$	& $[388  , 682]$ &	$17$	& $[3417.93  , 7965.11]$ &	$18$  \\\hline
\rowcolor{gray!25}ACE	& $[5  , 5]$ &	$13$	& $[37.42  , 62.13]$ &	$17$	& $[280  , 772]$ &	$18$	& $[2095.33  , 9592.70]$ &	$17$  \\\hline
TTM	& $[5  , 5]$ &	$13$	& $[39.69  , 46.85]$ &	$19$	& $[315  , 439]$ &	$19$	& $[2500.23  , 4113.50]$ &	$21$  \\\hline
ALG	& $[4  , 4]$ &	$19$	& $[31.81  , 43.27]$ &	$20$	& $[253  , 468]$ &	$20$	& $[2012.11  , 5062.19]$ &	$20$  \\\hline
AAL	& $[3  , 3]$ &	$21$	& $[23.75  , 39.72]$ &	$21$	& $[188  , 526]$ &	$21$	& $[1488.25  , 6964.95]$ &	$19$  \\\hline
ALI	& $[3  , 3]$ &	$21$	& $[23.17  , 34.07]$ &	$22$	& $[179  , 387]$ &	$22$	& $[1382.67  , 4395.48]$ &	$22$  \\\hline
BAL	& $[3  , 3]$ &	$21$	& $[24.80  , 29.55]$ &	$23$	& $[205  , 291]$ &	$23$	& $[1694.61  , 2866.02]$ &	$23$  \\\thickhline      
\end{tabular}
\begin{tablenotes}
      \tiny
      \item [a] {NUTS 3: ACE-Alentejo Central, ALI-Alentejo Litoral, ALG-Algarve, AAL-Alto Alentejo, AMI-Alto Minho, ATA-Alto T\^amega, AML-\'Area Metropolitana de Lisboa, AMP-\'Area Metropolitana do Porto, AVE-Ave, BAL-Baixo Alentejo, BBA-Beira Baixa, BSE-Beiras e Serra da Estrela, CAV-C\'avado, DOU-Douro, LTJ-Lez\'iria do Tejo, MTJ-M\'edio Tejo, OES-Oeste, RAV-Regi\~ao de Aveiro, RCO-Regi\~ao de Coimbra, RLE-Regi\~ao de Leiria, TES-T\^amega e Sousa, TTM-Terras de Tr\'as-os-Montes, VDL-Viseu D\~ao Laf\~oes.}
    \end{tablenotes}
\end{threeparttable}
\end{table}

\newpage

\paragraph{Interval-Weighted Flow Centrality measures: Betweenness (IWFB) and Closeness (IWFC)}

In Table~\ref{chp7_Results_IWD_Betweenness_table_CommutersNetwork} are shown the \textit{flow centrality} measures for the 23 NUTS 3, ranked in descending order according to the \textit{flow betweenness} centrality interval score. Regarding the \textit{Interval-Weighted Flow Betweenness} (IWFB), the five most central NUTS 3 with the highest max-flow between all pairs of NUTS 3, that depends on them are:  AML (Lisbon Metropolitan Area) $[15609,101172]$\footnote{This results means that of the total \textit{max-flow} between all pairs of NUTS 3 $[68404,323339]$, where AML is neither a source or a sink, $[15609,101172]$ must pass through AML.}, followed by AMP (Porto Metropolitan Area) $[7348,95972]$, RCO (Coimbra Region) $[5306,55549]$, RAV (Aveiro Region) $[2329,39281]$ and RLE (Leiria Region) $[2805,35233]$. These are the areas of the most important cities (Lisbon and Porto), and then 3 regions in the center of the country, through which important commuter flows must pass.

Nevertheless, regarding the \textit{Interval-Weighted Flow Closeness} (IWFC) centrality measure, the five most central NUTS 3 with the highest max-flow between them and the remaining NUTS 3 in the IWFC ranking (Table~\ref{chp7_Results_IWD_Betweenness_table_CommutersNetwork}) are: AMP (Porto Metropolitan Area) $[8645,55083]$ ranks 1\textsuperscript{st}, AVE (Ave Region) $[8020,55083]$, CAV (C\'avado Region) $[8048,54530]$, TES (T\^amega e Sousa Region) $[7154,51528]$ and AML (Lisbon Metropolitan Area) $[8645,49521]$. We find again the areas of Lisbon and Porto, as expected, plus three areas in the North of the country, known for being densely populated, therefore responsible for a greater flow of commuters between these regions.

It is noteworthy that AMP (Porto Metropolitan Area) becomes the most central region in the IWFC while in the IWFB it ranks 2\textsuperscript{nd}, AVE region ranks 2\textsuperscript{nd} in the IWFC while in the IWFB it ranks 10\textsuperscript{th}, CAV region ranks 3\textsuperscript{nd} in the IWFC while in the IWFB it ranks 6\textsuperscript{th}, TES region ranks 4\textsuperscript{th} in the IWFC while in the IWFB it ranks 14\textsuperscript{th}, whereas AML becomes 5\textsuperscript{th} in the IWFC while in the IWFB it ranks 1\textsuperscript{st}.

These differences between the IWFB and the IWFC centrality measures rankings are to be expected because the former measures the regions intermediary ability to communicate with other regions, high for AMP and AML, RCO, RAV, RLE, and the latter identifies the ``communication power'' of one region and the rest of the interval-weighted network, which is high for AMP, AML, AVE, CAV and  TES.

\begin{table}[h!]
\centering
\begin{threeparttable}
\caption{Flow centrality measures for the Interval-Weighted Commuters network (NUTS 3 ranked in descending order of \textit{interval rank} for IWFB).}
\label{chp7_Results_IWD_Betweenness_table_CommutersNetwork}
\renewcommand{\arraystretch}{.8}
\setlength{\tabcolsep}{2.1pt}
\fontsize{8}{10}\selectfont
\begin{tabular}{M{1.5cm}?M{2.25cm}|M{2cm}?M{2.5cm}?M{1.5cm}?M{2.25cm}|M{1.5cm}}
\thickhline
			& \multicolumn{2}{c?}{\textit{max-flow} (all pairs)\tnote{b}} & \multicolumn{2}{c?}{\textbf{Flow Betweenness}} & \multicolumn{2}{c}{\textbf{Flow Closeness}} \\ \hhline{~------}
NUTS 3\tnote{a}				  & \textit{max-flow} & \textit{max-flow} rank & \textbf{IWFB}\textsuperscript{c}    & IWFB rank      & \textbf{IWFC}\textsuperscript{d}        & IWFC rank  \\ \thickhline 
AML	& $[68404 , 323339]$ & $  19  $& $[15609 , 101172]$ & $	1 $	& $[8645 , 49521]$ & $	5  $ \\\hline
AMP	& $[68404 , 317777]$ & $	23	$& $[7348  , 95972]$ & $	2	$& $[8645  , 55083]$ & $	1  $ \\\hline
RCO	& $[68637 , 325757]$ & $	16	$& $[5306  , 55549]$ & $	3	$& $[8412  , 47103]$ & $	8  $ \\\hline
RAV	& $[69859 , 325331]$ & $	15	$& $[2329  , 39281]$ & $	4	$& $[7190  , 47529]$ & $	9  $ \\\hline
RLE	& $[69885 , 324177]$ & $	17	$& $[2805  , 35233]$ & $	5	$& $[7164  , 48683]$ & $	7  $ \\\hline
CAV	& $[69001 , 318330]$ & $	21	$& $[3481  , 27263]$ & $	6	$& $[8048  , 54530]$ & $	3  $ \\\hline
MTJ	& $[69663 , 327879]$ & $	14	$& $[4345  , 23542]$ & $	7	$& $[7386  , 44981]$ & $	10 $ \\\hline
DOU	& $[68887 , 347347]$ & $	11	$& $[4711  , 22248]$ & $	8	$& $[8162  , 25513]$ & $	13 $ \\\hline
VDL	& $[69429 , 350875]$ & $	10	$& $[1932  , 19480]$ & $	9	$& $[7620  , 21985]$ & $	14 $ \\\hline
AVE	& $[69029 , 317777]$ & $	22	$& $[3671  , 14260]$ & $	10	$& $[8020 , 55083]$ & $	2  $ \\\hline
BSE	& $[71490 , 352435]$ & $	9	$& $[1357   , 15864]$ & $	11	$& $[5559 , 20425]$ & $	15 $ \\\hline
ACE	& $[71217 , 357715]$ & $	7	$& $[4286   , 9969]$ & $	12	$& $[5832  , 15145]$ & $	17 $ \\\hline
OES	& $[70225 , 323339]$ & $	18	$& $[2207  , 10424]$ & $	13	$& $[6824 , 49521]$ & $	6  $ \\\hline
TES	& $[69895 , 321332]$ & $	20	$& $[1783  , 9860]$ & $	14	$& $[7154  , 51528]$ & $	4  $ \\\hline
LTJ	& $[71422 , 327051]$ & $	13	$& $[2656  , 8175]$ & $	15	$& $[5627  , 45809]$ & $	11 $ \\\hline
BBA	& $[70441 , 356215]$ & $	8	$& $[1582   , 8196]$ & $	16	$& $[6608  , 16645]$ & $	16 $ \\\hline
ALG	& $[71670 , 362851]$ & $	4	$& $[2822   , 6668]$ & $	17	$& $[5379  , 10009]$ & $	20 $ \\\hline
ATA	& $[69867 , 359155]$ & $	6	$& $[1886   , 6097]$ & $	18	$& $[7182  , 13705]$ & $	18 $ \\\hline
AAL	& $[72922 , 361807]$ & $	3	$& $[1607   , 6235]$ & $	19	$& $[4127  , 11053]$ & $	21 $ \\\hline
AMI	& $[71262 , 335269]$ & $	12	$& $[1190  , 5380]$ & $	20	$& $[5787  , 37591]$ & $	12 $ \\\hline
ALI	& $[73111 , 364442]$ & $	2	$& $[2419   , 3782]$ & $	21	$& $[3938  , 8418]$ & $	22  $ \\\hline
TTM	& $[70727 , 363402]$ & $	5	$& $[2045   , 3950]$ & $	22	$& $[6322  , 9458]$ & $	19  $ \\\hline
BAL	& $[72582 , 366458]$ & $	1	$& $[2256   , 3542]$ & $	23	$& $[4467  , 6402]$ & $	23$  \\\thickhline 
\end{tabular}
\begin{tablenotes}
      \tiny
	\item [a] {NUTS 3: ACE-Alentejo Central, ALI-Alentejo Litoral, ALG-Algarve, AAL-Alto Alentejo, AMI-Alto Minho, ATA-Alto T\^amega, AML-\'Area Metropolitana de Lisboa, AMP-\'Area Metropolitana do Porto, AVE-Ave, BAL-Baixo Alentejo, BBA-Beira Baixa, BSE-Beiras e Serra da Estrela, CAV-C\'avado, DOU-Douro, LTJ-Lez\'iria do Tejo, MTJ-M\'edio Tejo, OES-Oeste, RAV-Regi\~ao de Aveiro, RCO-Regi\~ao de Coimbra, RLE-Regi\~ao de Leiria, TES-T\^amega e Sousa, TTM-Terras de Tr\'as-os-Montes, VDL-Viseu D\~ao Laf\~oes.}      
      \item [b] {\textit{max-flow} between all pairs of vertices, where the vertex $v_i$ is neither a source or a sink; \textsuperscript{c} Interval--Weighted Flow Betweenness centrality; \textsuperscript{d} Interval--Weighted Flow Betweenness centrality.}
\end{tablenotes}
\end{threeparttable}
\end{table}

\subsection{Network of Trade transactions between 28 European countries}
\label{applied_exemple_2}

The construction of the `Interval-Weighted Trade Network (IWTN)'' of annual merchandise trade was done by aggregating the observations from the data corresponding to the values from $i\to j$ and from $j\to i$, between the 28 selected countries from 2003 to 2015~\citep{UNCTAD:2016a}\footnote{According to various authors, the \textit{flows} of annual merchandise trade can be conceived as a network \citep{Barigozzi:2011bd,Traag:2014tj,Barbosa:2018ki}.}, a \textit{temporal} aggregation~\citep{Brito:2014es}.
Therefore, each \textit{vertex} of the IWTN corresponds to one of the 28 European countries and the \textit{edges} represent intervals varying between the \textit{minimum} and \textit{maximum} exports (in thousands of US dollars) among those countries\footnote{Analogously to the procedure adopted for the IWCN in Section~\ref{applied_exemple_1} (see Figures~\ref{chp7_fig:directed_to_undirected_edges}a and \ref{chp7_fig:directed_to_undirected_edges}b), where the elements $o^I_{ij}$ of the symmetric interval-weighted adjacency matrix, $O^I$, denote the maximum variability of the \textit{bi-directional} flows $ij$ and $ji$ between the countries $i$ and $j$ (Figure\ref{chp7_fig:directed_to_undirected_edges}b): \linebreak$o^I_{ij}=\big[\min\{\underline{o}'_{ij},\underline{o}''_{ji}\},\max\{\overline{o}'_{ij},\overline{o}''_{ji}\}\big]=\big[\underline{o}_{ij},\overline{o}_{ij}\big]$.}.
\pagebreak

\begin{figure}[h!]
    \centering
    \begin{subfigure}[t]{0.60\linewidth}
        \includegraphics[width=1\linewidth,clip, trim={0cm 7cm 0cm 9.1cm}]{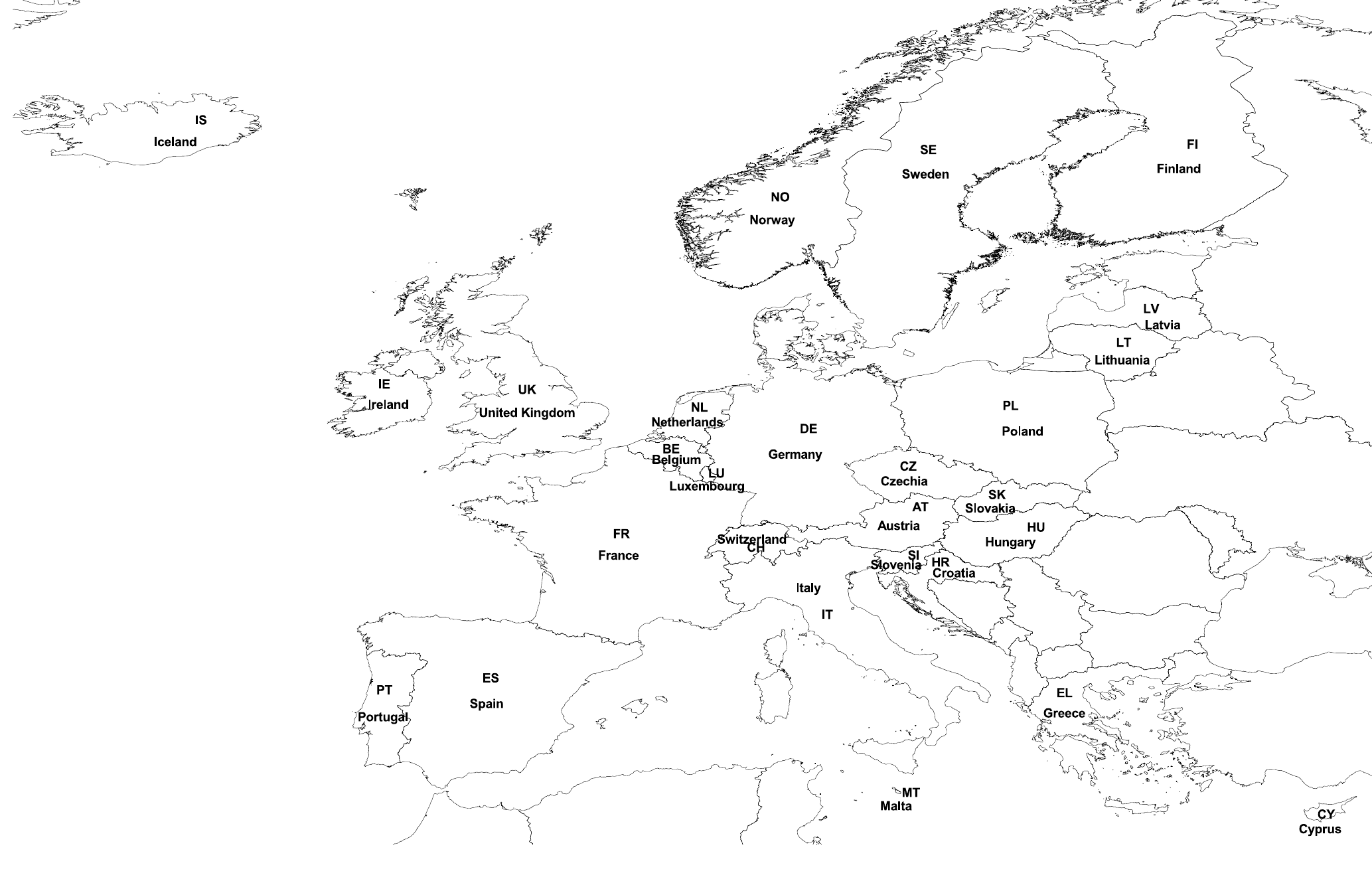}
        \caption{}
        \label{chp7_fig:Map_EU}
    \end{subfigure}%
    \begin{subfigure}[t]{0.40\linewidth}
            \includegraphics[width=1\linewidth,clip, trim={0cm 3cm 0cm 3cm}]{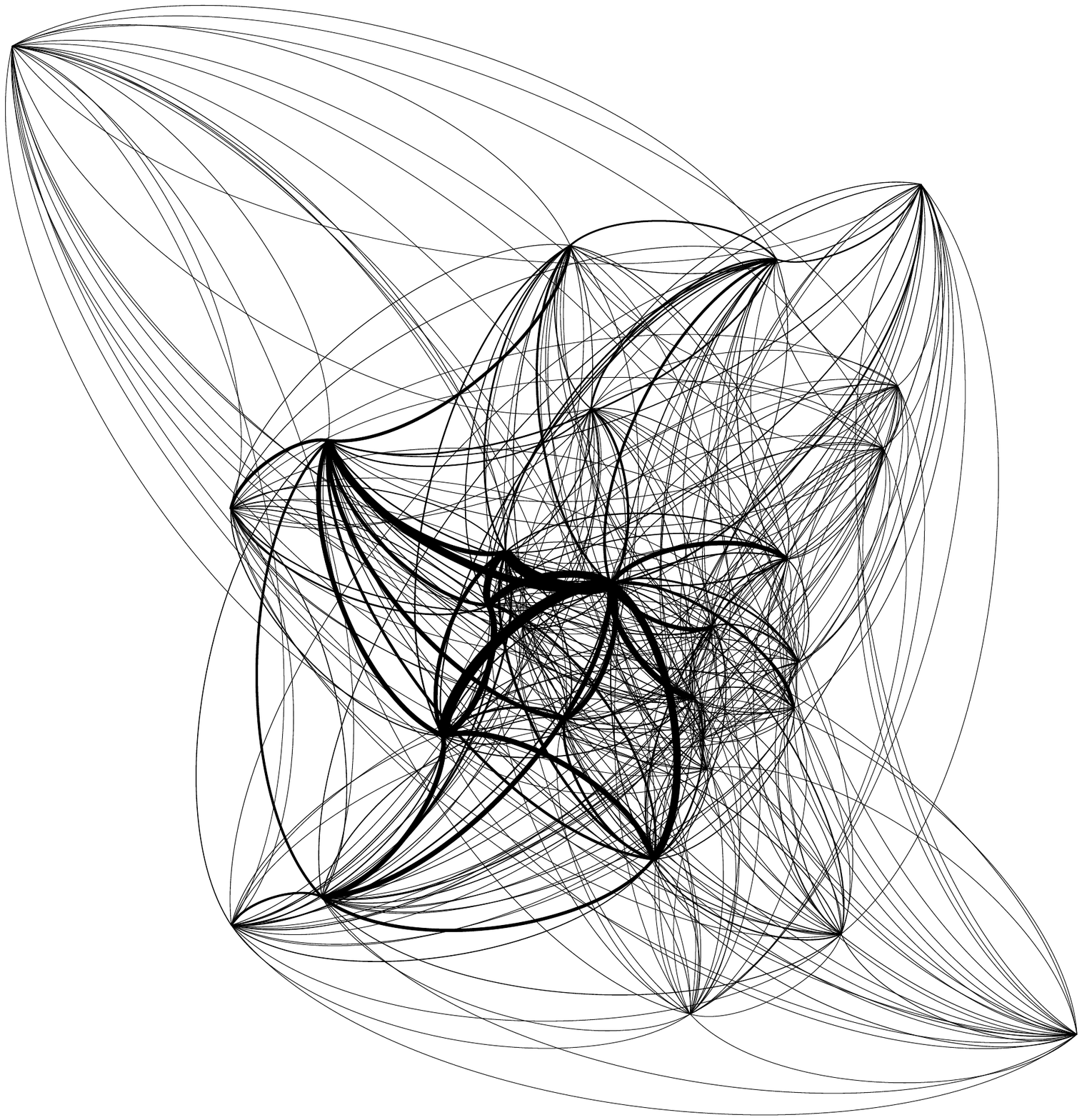}
        \caption{}
        \label{chp7_fig:Map_EUNet}
    \end{subfigure}
    \caption[Geographic representation of the 28 European countries (Trade network and the correspondent weighted network]{(a) Geographic representation of the European countries belonging to the ``Trade network'', and (b) Topologic representation of the weighted ``Trade network'' (weighted by the intervals midpoint).}
    \label{chp7_fig:Map_EUandNet}
\end{figure}

Figure~\ref{chp7_fig:Map_EUandNet} shows the geographical distribution of the European countries belonging to the ``Trade network'' (Figure\ref{chp7_fig:Map_EUandNet}a, and the corresponding network (Figure\ref{chp7_fig:Map_EUandNet}b)\footnote{For the sake of visualization, we chose not to represent the intervals on the network edges, such as it is depicted in Figure~\ref{chp7_fig:directed_to_undirected_edges}d.}. This is a \textit{complete} network (all vertices are connected between each other)  which has 28 vertices and 378 edges and is therefore considered a small network in size but with high density ($\text{graph density}=1.0$, $\text{diameter}=1$, $\text{average degree}=27$).

\subsubsection{Results -- interval-weighted trade network (IWTN)}
\label{chp7_Results_Centrality_Measures_TRADE}

\paragraph{Interval-Weighted Degree Centrality (IWD)}

Table~\ref{chp7_Results_IWD_Degree_table_TRADE} below shows the \textit{degree centrality} interval score for different values of the tuning parameter $\alpha$, ranking in descending order the 28 countries according to the \textit{degree centrality} accounting only for the weight of the edges $\alpha=1$\footnote{Highlighted in gray are the cases where there was a shift in the interval rank classification with the change of the value of $\alpha$.}. Since this is a \textit{complete network}, the number of edges attached to each one of the 28 countries analysed is the same (27 in this case), causing that shifting the tuning parameter $\alpha$ for the benchmark values $0.5, 1$, and $1.5$, does not cause significant changes in degree rankings. As it might be expected, the major European economies appear as the most central ones: DE (Germany), FR (France), UK (United Kingdom), NL (Netherlands), BE (Belgium), IT (Italy) and ES (Spain). 

\begin{landscape}
\begin{table}[h!]
\begin{threeparttable}
\caption{Degree centrality for the Interval-Weighted Trade network (Countries ranked in descending order of \textit{interval rank} for $\alpha=1$).}
\label{chp7_Results_IWD_Degree_table_TRADE}
\centering
\renewcommand{\arraystretch}{1}
\setlength{\tabcolsep}{2.1pt}
\fontsize{7.5}{10}\selectfont
\begin{tabular}{M{1.5cm}|M{1.5cm}|M{1cm}|M{3cm}|M{1cm}|M{3.5cm}|M{1cm}|M{4.5cm}|M{1cm}}
\thickhline
				 & \multicolumn{8}{c}{Tuning parameter $(\alpha)$} \\\cline{2-9} 
                        & \multicolumn{2}{c|}{$\alpha=0$}        & \multicolumn{2}{c|}{$\alpha=0.5$}       & \multicolumn{2}{c|}{$\alpha=1$}        & \multicolumn{2}{c}{$\alpha=1.5$}         \\\cline{2-9}
\multirow{2}{*}{Countries\tnote{a}}   & \makecell{Degree\\interval} & \makecell{Interval\\rank} & \makecell{Degree\\interval} & \makecell{Interval\\rank} & \makecell{Degree\\interval} & \makecell{Interval\\rank} & \makecell{Degree\\interval} & \makecell{Interval\\rank} \\\thickhline  
DE	&$[27,27]$&$	1$	&$[102326.85,175664.12]$&$	1$	&$[387806824.1,1142884585]$&$	1$	&$[1.469743e+12,7.435697e+12]$&$	1$\\\hline
FR	&$[27,27]$&$	1$	&$[78465.04,124536.28]$&$	2$	&$[228028235.1,574417976]$&$	2$	&$[6.626757e+11,2.649477e+12]$&$	2$\\\hline
UK	&$[27,27]$&$	1$	&$[68417.45,121986.45]$&$	3$	&$[173368436.3,551136848]$&$	3$	&$[4.393121e+11,2.490046e+12]$&$	3$\\\hline
NL	&$[27,27]$&$	1$	&$[64452.68,122444.18]$&$	4$	&$[153857308.5,555280599]$&$	4$	&$[3.672783e+11,2.518181e+12]$&$	4$\\\hline
BE	&$[27,27]$&$	1$	&$[62336.59,109086.05]$&$	5$	&$[143920368.4,440732048]$&$	5$	&$[3.322779e+11,1.780656e+12]$&$	5$\\\hline
IT	&$[27,27]$&$	1$	&$[65458.24,105143.01]$&$	6$	&$[158695591.1,409446369]$&$	6$	&$[3.847383e+11,1.594460e+12]$&$	6$\\\hline
ES	&$[27,27]$&$	1$	&$[53541.28,89445.76]$&$	7$	&$[106172922.1,296316422]$&$	7$	&$[2.105420e+11,9.816388e+11]$&$	7$\\\hline
CH	&$[27,27]$&$	1$	&$[40808.66,82191.47]$&$	8$	&$[61679515.7,250201410]$&$	8$	&$[9.322439e+10,7.616453e+11]$&$	8$\\\hline
AT	&$[27,27]$&$	1$	&$[40815.39,68822.90]$&$	9$	&$[61699855.2,175429336]$&$	9$	&$[9.327051e+10,4.471688e+11]$&$	10$\\\hline
PL	&$[27,27]$&$	1$	&$[33190.64,71855.75]$&$	10$	&$[40800682.2,191231455]$&$	10$	&$[5.015558e+10,5.089289e+11]$&$	9$\\\hline
SE	&$[27,27]$&$	1$	&$[39661.81,64544.17]$&$	11$	&$[58261436.6,154294441]$&$	11$	&$[8.558347e+10,3.688447e+11]$&$	11$\\\hline
\rowcolor{gray!25}NO	&$[27,27]$&$	1$	&$[25755.63,67581.50]$&$	13$	&$[24568609.2,169157757]$&$	12$	&$[2.343630e+10,4.234050e+11]$&$	12$\\\hline
\rowcolor{gray!25}CZ	&$[27,27]$&$	1$	&$[32407.79,64297.41]$&$	12$	&$[38898697.0,153116925]$&$	13$	&$[4.668966e+10,3.646304e+11]$&$	13$\\\hline
\rowcolor{gray!25}IE	&$[27,27]$&$	1$	&$[29851.83,53765.41]$&$	15$	&$[33004874.4,107063662]$&$	14$	&$[3.649096e+10,2.131971e+11]$&$	14$\\\hline
\rowcolor{gray!25}DK	&$[27,27]$&$	1$	&$[32359.85,52121.61]$&$	14$	&$[38783706.3,100617119]$&$	15$	&$[4.648278e+10,1.942343e+11]$&$	15$\\\hline
HU	&$[27,27]$&$	1$	&$[29042.88,49959.48]$&$	16$	&$[31240334.1,92442580]$&$	16$	&$[3.360405e+10,1.710512e+11]$&$	16$\\\hline
\rowcolor{gray!25}SK	&$[27,27]$&$	1$	&$[20855.16,46253.34]$&$	19$	&$[16108809.1,79235966]$&$	17$	&$[1.244266e+10,1.357381e+11]$&$	17$\\\hline
PT	&$[27,27]$&$	1$	&$[25589.03,43750.86]$&$	18$	&$[24251801.6,70893996]$&$	18$	&$[2.298445e+10,1.148768e+11]$&$	18$\\\hline
\rowcolor{gray!25}FI	&$[27,27]$&$	1$	&$[26188.34,43348.28]$&$	17$	&$[25401081.1,69595323]$&$	19$	&$[2.463749e+10,1.117347e+11]$&$	19$\\\hline
EL	&$[27,27]$&$	1$	&$[13783.88,39116.17]$&$	20$	&$[7036862.9,56669429]$&$	20$	&$[3.592417e+09,8.209966e+10]$&$	20$\\\hline
LU	&$[27,27]$&$	1$	&$[15505.02,30508.99]$&$	21$	&$[8903912.7,34474010]$&$	21$	&$[5.113161e+09,3.895434e+10]$&$	21$\\\hline
SI	&$[27,27]$&$	1$	&$[15318.91,28393.76]$&$	22$	&$[8691444.8,29859460]$&$	22$	&$[4.931239e+09,3.140082e+10]$&$	22$\\\hline
\rowcolor{gray!25}LT	&$[27,27]$&$	1$	&$[9470.52,25318.00]$&$	24$	&$[3321876.1,23740782]$&$	23$	&$[1.165181e+09,2.226182e+10]$&$	23$\\\hline
\rowcolor{gray!25}HR	&$[27,27]$&$	1$	&$[10479.74,24328.75]$&$	23$	&$[4067588.9,21921777]$&$	24$	&$[1.578788e+09,1.975294e+10]$&$	24$\\\hline
LV	&$[27,27]$&$	1$	&$[6983.80,20195.38]$&$	25$	&$[1806423.4,15105675]$&$	25$	&$[4.672480e+08,1.129870e+10]$&$	25$\\\hline
CY	&$[27,27]$&$	1$	&$[2968.57,17697.96]$&$	26$	&$[326386.2,11600664]$&$	26$	&$[3.588524e+07,7.604005e+09]$&$	26$\\\hline
IS	&$[27,27]$&$	1$	&$[5444.54,14558.85]$&$	27$	&$[1097891.1,7850370]$&$	27$	&$[2.213895e+08,4.233049e+09]$&$	28$\\\hline
MT	&$[27,27]$&$	1$	&$[3984.12,14765.58]$&$	28$	&$[587897.3,8074900]$&$	28$	&$[8.675016e+07,4.415947e+09]$&$	27$\\\thickhline      
\end{tabular}
\begin{tablenotes}
      \tiny
	\item [a] {Countries: AT-Austria, BE-Belgium, HR-Croatia, CY-Cyprus, CZ-Czech Republic, DK-Denmark, FI-Finland, FR, France, DE-Germany, EL-Greece, HU-Hungary, IS-Iceland, IE-Ireland, IT-Italy, LV-Latvia, LT, Lithuania, LU-Luxembourg, MT-Malta, NL-Netherlands, NO-Norway, PL-Poland, PT-Portugal, SK-Slovakia, SI-Slovenia, ES-Spain, SE-Sweden, CH-Switzerland, UK-United Kingdom.}      
\end{tablenotes}
\end{threeparttable}
\end{table}
\end{landscape}

\paragraph{Interval-Weighted Flow Centrality measures: Betweenness (IWFB) and Closeness (IWFC)}

Table~\ref{chp7_Results_IWD_Betweenness_table_TRADE} shows the \textit{flow centrality} measures for the 28 European countries analysed, ranked in descending order according to the \textit{flow betweenness} centrality interval score. Regarding the \textit{Interval-Weighted Flow Betweenness} (IWFB), DE (Germany) is the most central European country, i.e., the country through which most flows must pass, followed by UK (United Kingdom), FR (France). NL (Netherlands), IT (Italy), BE (Belgium) and ES (Spain). 

Nevertheless, regarding the \textit{Interval-Weighted Flow Closeness} (IWFC) centrality measure, FR (France) becomes the most central European country, i.e., the country that centralizes most annual merchandise trade in the considered years (Table~\ref{chp7_Results_IWD_Betweenness_table_CommutersNetwork}).

It is noteworthy that SE (Sweden) ranks 11\textsuperscript{th} in the IWFC while in the IWFB it ranks 8\textsuperscript{th}, putting in evidence its intermediation role. On the contrary, CH (Switzerland) ranks 8\textsuperscript{th} in the IWFC whereas in the IWFB it ranks 10\textsuperscript{th}.

\begin{landscape}
\begin{table}[h!]
\centering
\begin{threeparttable}
\caption{Flow centrality measures for the Interval--Weighted Trade network (Countries ranked in descending order of \textit{interval rank} for IWFB).}
\label{chp7_Results_IWD_Betweenness_table_TRADE}
\centering
\renewcommand{\arraystretch}{1}
\setlength{\tabcolsep}{2.1pt}
\fontsize{7.5}{10}\selectfont
\begin{tabular}{M{1.5cm}?M{4cm}|M{2cm}?M{4cm}?M{1.5cm}?M{4cm}|M{1.5cm}}
\thickhline
			& \multicolumn{2}{c?}{\textit{max--flow} (all pairs)\tnote{b}} & \multicolumn{2}{c?}{\textbf{Flow Betweenness}} & \multicolumn{2}{c}{\textbf{Flow Closeness}} \\ \hhline{~------}
Countries\tnote{a}				  & \textit{max--flow} & \textit{max--flow} rank & \textbf{IWFB}\textsuperscript{c}    & IWFB rank      & \textbf{IWFC}\textsuperscript{d}        & IWFC rank  \\ \thickhline 
DE	&$[7413003807, 26769887480]$&$	27$	&$[2185413096, 6566281708]$&$	1$	&$[1453915546, 4649907299]$&$	1$\\\hline
UK	&$[7468330607, 26797312358]$&$	26	$&$[838121364, 2978511281]$&$	2$	&$[1398588745, 4622482420]$&$	3$\\\hline
FR	&$[7413003807, 26769887480]$&$	27	$&$[881547435, 2328131952]$&$	3$	&$[1453915546, 4649907299]$&$	1$\\\hline
NL	&$[7510857142, 26789024857]$&$	25$	&$[557453333, 2159124156]$&$	4$	&$[1356062210, 4630769922]$&$	4$\\\hline
IT	&$[7496342294, 27253669475]$&$	23	$&$[633270908, 2072687573]$&$	5$	&$[1370577058, 4166125304]$&$	6$\\\hline
BE	&$[7550604902, 27128526758]$&$	24$	&$[453239198, 1752227012]$&$	6$	&$[1316314450, 4291268020]$&$	5$\\\hline
ES	&$[7739342134, 27819319211]$&$	22$	&$[401228536, 1484669303]$&$	7$	&$[1127577219, 3600475569]$&$	7$\\\hline
SE	&$[8033667545, 28840293295]$&$	18$	&$[383023300, 1051732575]$&$	8$	&$[833251808, 2579501489]$&$	11$\\\hline
PL	&$[8190814334, 28508798969]$&$	19$	&$[157392715, 1013176935]$&$	9$	&$[676105018, 2910995812]$&$	10$\\\hline
CH	&$[8006322912, 28096009281]$&$	21$	&$[179713794, 894364013]$&$	10$	&$[860596441, 3323785497]$&$	8$\\\hline
AT	&$[8006180535, 28635215917]$&$	20$	&$[243111799, 784248460]$&$	11$ &$[860738817, 2784578860]$&$	9$\\\hline
CZ	&$[8209834186, 28853245967]$&$	17$	&$[163521022, 799346141]$&$	12$	&$[657085166, 2566548813]$&$	12$\\\hline
NO	&$[8397620711, 28691660132]$&$	16$	&$[143183388, 799696175]$&$	13$	&$[469298641, 2728134649]$&$	13$\\\hline
DK	&$[8211099084, 29489690179]$&$	14$	&$[206979675, 633196218]$&$	14$	&$[655820269, 1930104598]$&$	15$\\\hline
HU	&$[8303384091, 29604133730]$&$	13$	&$[119391634, 522797634]$&$	15$	&$[563535262, 1815661052]$&$	16$\\\hline
SK	&$[8541120505, 29802232940]$&$	10$	&$[87072347, 546677594]$&$	16	$&$[325798847, 1617561842]$&$	19$\\\hline
FI	&$[8385133632, 29957781899]$&$	11$	&$[134826401, 453203323]$&$	17$	&$[481785719, 1462012881]$&$	18$\\\hline
IE	&$[8280445067, 29405885128]$&$	15$	&$[98708914, 465212077]$&$	18	$&$[586474286, 2013909657]$&$	14$\\\hline
PT	&$[8402689632, 29935704453]$&$	12$	&$[92016005, 328147696]$&$	19$	&$[464229719, 1484090322]$&$	17$\\\hline
EL	&$[8707937169, 30190447994]$&$	9$	&$[27673872, 296713444]$&$	20	$&$[158982184, 1229346789]$&$	20$\\\hline
SI	&$[8674845530, 30704451956]$&$	7$	&$[43474161, 211867658]$&$	21$	&$[192073822, 715342828]$&$	22$\\\hline
LT	&$[8786697603, 30832944183]$&$	6$	&$[21970780, 204033734]$&$	22$	&$[80221748, 586850590]$&$	23$\\\hline
HR	&$[8770291922, 30872962304]$&$	5$	&$[24009255, 184513589]$&$	23$	&$[96627429, 546832480]$&$	24$\\\hline
LV	&$[8821553016, 31029732634]$&$	4$	&$[14592275, 160007089]$&$	24$	&$[45366336, 390062134]$&$	25$\\\hline
LU	&$[8670808641, 30612160946]$&$	8$	&$[23517667, 135821723]$&$	25$	&$[196110712, 807633828]$&$	21$\\\hline
CY	&$[8858106924, 31113852903]$&$	3$	&$[2678597, 98121070]$&$	26$	&$[8812427, 305941870]$&$	26$\\\hline
IS	&$[8838557792, 31207834800]$&$	2$	&$[5273107, 54270679]$&$	27$	&$[28361561, 211959990]$&$	27$\\\hline
MT	&$[8851307637, 31201997000]$&$	1$	&$[1342859, 50523695]$&$	28$	&$[15611716, 217797770]$&$	28$\\\hline
\end{tabular}
\begin{tablenotes}
      \tiny
	\item [a] {Countries: AT-Austria, BE-Belgium, HR-Croatia, CY-Cyprus, CZ-Czech Republic, DK-Denmark, FI-Finland, FR, France, DE-Germany, EL-Greece, HU-Hungary, IS-Iceland, IE-Ireland, IT-Italy, LV-Latvia, LT, Lithuania, LU-Luxembourg, MT-Malta, NL-Netherlands, NO-Norway, PL-Poland, PT-Portugal, SK-Slovakia, SI-Slovenia, ES-Spain, SE-Sweden, CH-Switzerland, UK-United Kingdom.}      
      \item [b] {\textit{max--flow} between between all pairs of vertices, where the vertex $v_i$ is neither a source or a sink; \textsuperscript{c} Interval--Weighted Flow Betweenness centrality; \textsuperscript{d} Interval--Weighted Flow Closeness centrality.}
\end{tablenotes}
\end{threeparttable}
\end{table}
\end{landscape}

\section{Concluding remarks}
\label{conclusion}

In recent years, the term ``Big Data'' emerged, and new approaches arise to deal with large amounts of information, including the possibility to aggregate data to provide more manageably-sized datasets. These new approaches may consist of considering aggregated (e.g. interval) data, keeping the information on the intrinsic variability in order to capture the original dispersion of the data. However, the definitions of basic statistical notions do not apply automatically in this aggregated data, and well-established properties are no longer straightforward. 
Furthermore, when we use such aggregated data on complex structures, such as network data, the situation may become harder do handle. Therefore, to apply statistical and multivariate data analysis techniques to interval data in network structures requires proper consideration and often the design of new approaches and appropriate techniques. 
In this paper we provide a novel contribution to network science in that we use aggregate interval data to describe the weights of networks' edges, giving rise to the concept of Interval-Weighted Networks (IWN). We start by generalizing the three classical centrality measures, degree, closeness and betweenness, for the general case of IWN, with a triple motivation: firstly, we try to establish a benchmark for these measures when using intervals defined by the minimum and maximum observed weights on the edges of the IWN; secondly, extend the degree centrality based on~\citet{Opsahl:2010in} concept of a tuning parameter to give relevance either to tie weights or number of ties alternatively; and thirdly, generalize closeness and betweenness based on network flows, where with each edge is assigned a flow which maximizes the total flow between a pair of vertices and using Ford and Fulkerson's max-flow method~\citep{Ford:1956vc,Freeman:1991un}.

The experiments carried out on an artificial network and on two real-world networks (IWCN and IWTN) have shown that, for the Interval-Weighted Degree (IWD),  as expected, the variation of the tuning parameter $\alpha$ to give relevance either to tie weights or number of ties alternatively~\citep{Opsahl:2010in} affects the ranking centrality of the vertices. In the IWCN it changes the topological importance of some NUTS 3 as an attraction point (center). In the IWTN, as the number of connections is the same for all vertices (the 28 European countries), i.e., is a complete network, the change is residual.

Similarly, it has been found that the use of intervals has made it possible to capture a variation in the flow betweenness (IWFB) and flow closeness (IWFC), in terms of all paths connecting pairs of vertices, and not based only on geodesic paths.

The joint use of the two measures allowed putting in evidence the regions/countries that play an important intermediation role, distinguishing them from the regions/countries that register a high flow with the rest of the network.

\paragraph{Acknowledgements:}
This work was financed by the Portuguese funding agency,\linebreak FCT - Funda\c{c}\~ao para a Ci\^encia e a Tecnologia, within project UIDB/50014/2020.
This research has also received funding from the European Union's Horizon 2020 research and innovation program ''FIN-TECH: A Financial supervision and Technology compliance training programme'' under the grant agreement No 825215 (Topic: ICT-35-2018, Type of action: CSA).

\renewcommand{\refname}{\normalsize References}
\bibliographystyle{chicago}
\bibliography{references}

\begin{thebibliography}{}

\bibitem[\protect\citeauthoryear{Ahuja, Magnanti, and Orlin}{Ahuja
  et~al.}{1993}]{Ahuja:1993uh}
Ahuja, R.~K., T.~L. Magnanti, and J.~B. Orlin (1993).
\newblock {\em {Network Flows}}.
\newblock Theory, algorithms, and applications. New Jersey: Prentice Hall.

\bibitem[\protect\citeauthoryear{Barabasi}{Barabasi}{2016}]{Barabasi:2016vs}
Barabasi, A.-L. (2016).
\newblock {\em {Network Science}}.
\newblock Cambridge University Press.

\bibitem[\protect\citeauthoryear{Barbosa, Barthelemy, Ghoshal, James,
  Lenormand, Louail, Menezes, Ramasco, Simini, and Tomasini}{Barbosa
  et~al.}{2018}]{Barbosa:2018ki}
Barbosa, H., M.~Barthelemy, G.~Ghoshal, C.~R. James, M.~Lenormand, T.~Louail,
  R.~Menezes, J.~J. Ramasco, F.~Simini, and M.~Tomasini (2018).
\newblock {Human mobility: Models and applications}.
\newblock {\em Physics Reports\/}~{\em 734}, 1--74.

\bibitem[\protect\citeauthoryear{Barigozzi, Fagiolo, and Mangioni}{Barigozzi
  et~al.}{2011}]{Barigozzi:2011bd}
Barigozzi, M., G.~Fagiolo, and G.~Mangioni (2011).
\newblock {Identifying the community structure of the international-trade
  multi-network}.
\newblock {\em Physica A: Statistical Mechanics and its Applications\/}~{\em
  390\/}(11), 2051--2066.

\bibitem[\protect\citeauthoryear{Barrat, Barthelemy, Pastor-Satorras, and
  Vespignani}{Barrat et~al.}{2004}]{2004PNAS..101.3747B}
Barrat, A., M.~Barthelemy, R.~Pastor-Satorras, and A.~Vespignani (2004).
\newblock {The architecture of complex weighted networks}.
\newblock {\em PNAS\/}~{\em 101\/}(11), 3747--3752.

\bibitem[\protect\citeauthoryear{Billard and Diday}{Billard and
  Diday}{2007}]{billard2006symbolic}
Billard, L. and E.~Diday (2007).
\newblock {\em {Symbolic Data Analysis: Conceptual Statistics and Data
  Mining}}.
\newblock Wiley Series in Computational Statistics. West Sussex, England:
  Wiley.

\bibitem[\protect\citeauthoryear{Bonacich}{Bonacich}{1972}]{Bonacich:1972dt}
Bonacich, P. (1972).
\newblock {Factoring and weighting approaches to status scores and clique
  identification}.
\newblock {\em The Journal of Mathematical Sociology\/}~{\em 2\/}(1), 113--120.

\bibitem[\protect\citeauthoryear{Bonacich}{Bonacich}{1987}]{Bonacich:1987up}
Bonacich, P. (1987).
\newblock {Power and centrality: A family of measures}.
\newblock {\em American journal of sociology\/}~{\em 92}, 1170--1182.

\bibitem[\protect\citeauthoryear{Borgatti}{Borgatti}{2005}]{Borgatti:2005je}
Borgatti, S.~P. (2005).
\newblock {Centrality and network flow}.
\newblock {\em Social Networks\/}~{\em 27\/}(1), 55--71.

\bibitem[\protect\citeauthoryear{Borgatti and Everett}{Borgatti and
  Everett}{2006}]{Borgatti:2006cf}
Borgatti, S.~P. and M.~G. Everett (2006).
\newblock {A Graph-theoretic perspective on centrality}.
\newblock {\em Social Networks\/}~{\em 28\/}(4), 466--484.

\bibitem[\protect\citeauthoryear{Bozhenyuk, Gerasimenko, Kacprzyk, and
  Rozenberg}{Bozhenyuk et~al.}{2017}]{Bozhenyuk:2017vd}
Bozhenyuk, A.~V., E.~M. Gerasimenko, J.~Kacprzyk, and I.~N. Rozenberg (2017).
\newblock {\em {Flows in Networks Under Fuzzy Conditions}}, Volume 346.
\newblock Studies in Fuzziness and Soft Computing.

\bibitem[\protect\citeauthoryear{Brandes}{Brandes}{2001}]{Brandes:2001wm}
Brandes, U. (2001).
\newblock {A Faster Algorithm for Betweenness Centrality}.
\newblock {\em Journal of mathematical sociology\/}~{\em 25\/}(2), 163--177.

\bibitem[\protect\citeauthoryear{Brandes}{Brandes}{2008}]{Brandes:2008gb}
Brandes, U. (2008).
\newblock {On variants of shortest-path betweenness centrality and their
  generic computation}.
\newblock {\em Social Networks\/}~{\em 30\/}(2), 136--145.

\bibitem[\protect\citeauthoryear{Brandes, Borgatti, and Freeman}{Brandes
  et~al.}{2016}]{Brandes:2016id}
Brandes, U., S.~P. Borgatti, and L.~C. Freeman (2016).
\newblock {Maintaining the duality of closeness and betweenness centrality}.
\newblock {\em Social Networks\/}~{\em 44}, 153--159.

\bibitem[\protect\citeauthoryear{Brandes and Fleischer}{Brandes and
  Fleischer}{2005}]{Brandes:2005ug}
Brandes, U. and D.~Fleischer (2005).
\newblock {Centrality measures based on current flow}.
\newblock In {\em Stacs 2005, Proceedings}, pp.\  533--544.

\bibitem[\protect\citeauthoryear{Brin and Page}{Brin and
  Page}{1998}]{Brin:1998vm}
Brin, S. and L.~Page (1998).
\newblock {The anatomy of a large-scale hypertextual Web search engine}.
\newblock {\em Computer Networks and Isdn Systems\/}~{\em 30\/}(1-7), 107--117.

\bibitem[\protect\citeauthoryear{Brito}{Brito}{2014}]{Brito:2014es}
Brito, P. (2014).
\newblock {Symbolic Data Analysis: another look at the interaction of Data
  Mining and Statistics}.
\newblock {\em Wiley Interdisciplinary Reviews: Data Mining and Knowledge
  Discovery\/}~{\em 4\/}(4), 281--295.

\bibitem[\protect\citeauthoryear{Cheng, Lee, Lim, and Zhu}{Cheng
  et~al.}{2015}]{Cheng:2015in}
Cheng, Y.-Y., R.~K.-W. Lee, E.-P. Lim, and F.~Zhu (2015).
\newblock {Measuring Centralities for Transportation Networks Beyond
  Structures}.
\newblock In P.~Kazienko and N.~V. Chawla (Eds.), {\em Applications of Social
  Media and Social Network Analysis}, Lecture Notes in Social Networks, pp.\
  23--39. Cham: Springer International Publishing.

\bibitem[\protect\citeauthoryear{Couso and Dubois}{Couso and
  Dubois}{2014}]{Couso:2014du}
Couso, I. and D.~Dubois (2014).
\newblock {Statistical reasoning with set-valued information: Ontic vs.
  epistemic views}.
\newblock {\em International Journal of Approximate Reasoning\/}~{\em 55\/}(7),
  1502--1518.

\bibitem[\protect\citeauthoryear{Dawood}{Dawood}{2011}]{Dawood:2011vh}
Dawood, H. (2011).
\newblock {\em {Theories of Interval Arithmetic}}.
\newblock Mathematical Foundations and Applications. LAP Lambert Academic
  Publishing.

\bibitem[\protect\citeauthoryear{De~Leo, Santoboni, Cerina, Mureddu, Secchi,
  and Chessa}{De~Leo et~al.}{2013}]{DeLeo:2013do}
De~Leo, V., G.~Santoboni, F.~Cerina, M.~Mureddu, L.~Secchi, and A.~Chessa
  (2013).
\newblock {Community core detection in transportation networks}.
\newblock {\em Physical Review E\/}~{\em 88\/}(4), 3.

\bibitem[\protect\citeauthoryear{De~Montis, Barthelemy, Chessa, and
  Vespignani}{De~Montis et~al.}{2007}]{DeMontis:2007iu}
De~Montis, A., M.~Barthelemy, A.~Chessa, and A.~Vespignani (2007).
\newblock {The Structure of Interurban Traffic: A Weighted Network Analysis}.
\newblock {\em Environment and Planning B: Planning and Design\/}~{\em
  34\/}(5), 905--924.

\bibitem[\protect\citeauthoryear{De~Montis, Caschili, and Chessa}{De~Montis
  et~al.}{2011}]{DeMontis:2011db}
De~Montis, A., S.~Caschili, and A.~Chessa (2011).
\newblock {Time evolution of complex networks: commuting systems in insular
  Italy}.
\newblock {\em Journal of Geographical Systems\/}~{\em 13\/}(1), 49--65.

\bibitem[\protect\citeauthoryear{De~Montis, Caschili, and Chessa}{De~Montis
  et~al.}{2013}]{DeMontis:2013ho}
De~Montis, A., S.~Caschili, and A.~Chessa (2013).
\newblock {Commuter networks and community detection: A method for planning sub
  regional areas}.
\newblock {\em The European Physical Journal Special Topics\/}~{\em 215\/}(1),
  75--91.

\bibitem[\protect\citeauthoryear{Dijkstra}{Dijkstra}{1959}]{Dijkstra:1959vb}
Dijkstra, E.~W. (1959).
\newblock {A note on two problems in connection with graphs}.
\newblock {\em Numer Math\/}~{\em 1}, 269--271.

\bibitem[\protect\citeauthoryear{Du, Gao, Hu, Mahadevan, and Deng}{Du
  et~al.}{2014}]{Du:2014hp}
Du, Y., C.~Gao, Y.~Hu, S.~Mahadevan, and Y.~Deng (2014).
\newblock {A new method of identifying influential nodes in complex networks
  based on TOPSIS}.
\newblock {\em Physica A: Statistical Mechanics and its Applications\/}~{\em
  399}, 57--69.

\bibitem[\protect\citeauthoryear{Eurostat}{Eurostat}{2016}]{Eurostat:2016a}
Eurostat (2016).
\newblock {Commission Regulation (EU) 2016/2066 of 21 November 2016 amending
  the annexes to Regulation (EC) No 1059/2003 of the European Parliament and of
  the Council on the establishment of a common classification of territorial
  units for statistics (NUTS)}.
\newblock Available online at:
  \url{https://ec.europa.eu/eurostat/web/nuts/background} (accessed:
  15.06.2017).

\bibitem[\protect\citeauthoryear{Ford and Fulkerson}{Ford and
  Fulkerson}{1956}]{Ford:1956vc}
Ford, L.~R. and D.~R. Fulkerson (1956).
\newblock {Maximal flow through a network}.
\newblock {\em Canadian journal of Mathematics\/}~{\em 8\/}(3), 399--404.

\bibitem[\protect\citeauthoryear{Ford and Fulkerson}{Ford and
  Fulkerson}{1957}]{Ford:1957vq}
Ford, L.~R. and D.~R. Fulkerson (1957).
\newblock {A simple algorithm for finding maximal network flows and an
  application to the Hitchcock problem}.
\newblock {\em Canada Journal of Mathematics\/}~{\em 9}, 210--218.

\bibitem[\protect\citeauthoryear{Ford and Fulkerson}{Ford and
  Fulkerson}{1962}]{Ford:26m8xm4j}
Ford, L.~R. and D.~R. Fulkerson (1962).
\newblock {\em {Flows in Networks}}.
\newblock NJ: Princeton University Press.

\bibitem[\protect\citeauthoryear{Freeman}{Freeman}{1977}]{Freeman:1977vn}
Freeman, L.~C. (1977).
\newblock {A set of measures of centrality based on betweenness}.
\newblock {\em Sociometry\/}, 35--41.

\bibitem[\protect\citeauthoryear{Freeman}{Freeman}{1979}]{Freeman:1979wx}
Freeman, L.~C. (1979).
\newblock {Centrality in social networks conceptual clarification}.
\newblock {\em Social Networks\/}~{\em 1\/}(3), 215--239.

\bibitem[\protect\citeauthoryear{Freeman, Borgatti, and White}{Freeman
  et~al.}{1991}]{Freeman:1991un}
Freeman, L.~C., S.~P. Borgatti, and D.~R. White (1991).
\newblock {Centrality in valued graphs: A measure of betweenness based on
  network flow}.
\newblock {\em Social Networks\/}~{\em 13\/}(2), 141--154.

\bibitem[\protect\citeauthoryear{Garas, Schweitzer, and Havlin}{Garas
  et~al.}{2012}]{Garas:2012em}
Garas, A., F.~Schweitzer, and S.~Havlin (2012).
\newblock {A k-shell decomposition method for weighted networks}.
\newblock {\em New Journal of Physics\/}~{\em 14\/}(8), 083030.

\bibitem[\protect\citeauthoryear{Ghalmane, El~Hassouni, Cherifi, and
  Cherifi}{Ghalmane et~al.}{2019}]{Ghalmane:2019ev}
Ghalmane, Z., M.~El~Hassouni, C.~Cherifi, and H.~Cherifi (2019).
\newblock {Centrality in modular networks}.
\newblock {\em EPJ Data Sci.\/}~{\em 8}, 1--27.

\bibitem[\protect\citeauthoryear{G{\'o}mez, Figueira, and
  Eus{\'e}bio}{G{\'o}mez et~al.}{2013}]{Gomez:2013ee}
G{\'o}mez, D., J.~R. Figueira, and A.~Eus{\'e}bio (2013).
\newblock {Modeling centrality measures in social network analysis using
  bi-criteria network flow optimization problems}.
\newblock {\em European Journal of Operational Research\/}~{\em 226\/}(2),
  354--365.

\bibitem[\protect\citeauthoryear{G{\'o}mez, Gonz{\'a}lez-Arang{\"u}ena, Manuel,
  Owen, del Pozo, and Tejada}{G{\'o}mez et~al.}{2003}]{Gomez:2003ud}
G{\'o}mez, D., E.~Gonz{\'a}lez-Arang{\"u}ena, C.~Manuel, G.~Owen, M.~del Pozo,
  and J.~Tejada (2003).
\newblock {Centrality and power in social networks - a game theoretic
  approach.}
\newblock {\em Mathematical Social Sciences\/}~{\em 46}, 27--54.

\bibitem[\protect\citeauthoryear{Granovetter}{Granovetter}{1973}]{Granovetter:1973wj}
Granovetter, M.~S. (1973).
\newblock {The strength of weak ties}.
\newblock {\em American journal of sociology\/}~{\em 78\/}(6), 1360--1380.

\bibitem[\protect\citeauthoryear{Grzegorzewski and {\'S}piewak}{Grzegorzewski
  and {\'S}piewak}{2017}]{Grzegorzewski:2016fq}
Grzegorzewski, P. and M.~{\'S}piewak (2017).
\newblock {The Sign Test for Interval-Valued Data}.
\newblock In {\em Soft Methods for Data Science. SMPS 2016. Advances in
  Intelligent Systems and Computing}, pp.\  269--276. Cham: Springer
  International Publishing.

\bibitem[\protect\citeauthoryear{Guerra and Stefanini}{Guerra and
  Stefanini}{2012}]{Guerra:2012hg}
Guerra, M.~L. and L.~Stefanini (2012).
\newblock {A comparison index for interval ordering based on generalized
  Hukuhara difference}.
\newblock {\em Soft Computing\/}~{\em 16\/}(11), 1931--1943.

\bibitem[\protect\citeauthoryear{Hossain}{Hossain}{2009}]{Hossain:2010vq}
Hossain, A. (2009).
\newblock {Most Reliable Route Method and Algorithm Based on Interval
  Possibilities for a Cyclic Network}.
\newblock {\em Cybernetics and Information Technologies\/}~{\em 9}, 81--92.

\bibitem[\protect\citeauthoryear{Hossain and Gatev}{Hossain and
  Gatev}{2010}]{Hossain:2010ur}
Hossain, A. and G.~Gatev (2010).
\newblock {Method and Algorithm for Interval Maximum Expected Flow in a
  Network}.
\newblock {\em Information technologies and control\/}~{\em 1}, 18--24.

\bibitem[\protect\citeauthoryear{Hu and Wang}{Hu and Wang}{2006}]{Hu:2006tq}
Hu, B.~Q. and S.~Wang (2006).
\newblock {A novel approach in uncertain programming part I: New arithmetic and
  order relation for interval numbers}.
\newblock {\em Journal of Industrial and Management Optimization\/}~{\em
  2\/}(4), 351--371.

\bibitem[\protect\citeauthoryear{Hu and Hu}{Hu and Hu}{2008}]{Hu:2008vt}
Hu, C. and P.~Hu (2008).
\newblock {Interval-Weighted Graphs and Flow Networks}.
\newblock In C.~Hu, R.~B. Kearfott, A.~d. Korvin, and V.~Kreinovich (Eds.),
  {\em Knowledge Processing with Interval and Soft Computing}, pp.\  1--16.
  London: Springer London.

\bibitem[\protect\citeauthoryear{Karmakar and Bhunia}{Karmakar and
  Bhunia}{2012}]{Karmakar:2012vm}
Karmakar, S. and A.~K. Bhunia (2012).
\newblock {A Comparative Study of Different Order Relations of Intervals.}
\newblock {\em Reliable Computing\/}~{\em 16}, 38--72.

\bibitem[\protect\citeauthoryear{Karmakar and Bhunia}{Karmakar and
  Bhunia}{2014}]{Karmakar:2014jo}
Karmakar, S. and A.~K. Bhunia (2014).
\newblock {Uncertain constrained optimization by interval-oriented algorithm}.
\newblock {\em Journal of the Operational Research Society\/}~{\em 65\/}(1),
  73--87.

\bibitem[\protect\citeauthoryear{Katz}{Katz}{1953}]{Katz:1953un}
Katz, L. (1953).
\newblock {A new status index derived from sociometric analysis}.
\newblock {\em Psychometrika\/}~{\em 18}, 39--43.

\bibitem[\protect\citeauthoryear{Lu, Chen, Ren, Zhang, Zhang, and Zhou}{Lu
  et~al.}{2016}]{Lu:2016bv}
Lu, L., D.~Chen, X.-L. Ren, Q.-M. Zhang, Y.-C. Zhang, and T.~Zhou (2016).
\newblock {Vital nodes identification in complex networks}.
\newblock {\em Physics Reports\/}~{\em 650}, 1--63.

\bibitem[\protect\citeauthoryear{Lu, Zhou, Zhang, and Stanley}{Lu
  et~al.}{2016}]{Lu:2016kl}
Lu, L., T.~Zhou, Q.-M. Zhang, and H.~E. Stanley (2016).
\newblock {The H-index of a network node and its relation to degree and
  coreness}.
\newblock {\em Nature Communications\/}~{\em 7}, 1--7.

\bibitem[\protect\citeauthoryear{Martin, Zhang, and Newman}{Martin
  et~al.}{2014}]{Martin:2014tr}
Martin, T., X.~Zhang, and M.~Newman (2014).
\newblock {Localization and centrality in networks}.
\newblock {\em Physical Review E\/}~{\em 90\/}(5), 052808.

\bibitem[\protect\citeauthoryear{Moore, Kearfott, and Cloud}{Moore
  et~al.}{2009}]{Moore:2009uc}
Moore, R.~E., R.~B. Kearfott, and M.~J. Cloud (2009).
\newblock {\em {Introduction to Interval Analysis}}.
\newblock Philadelphia: SIAM.

\bibitem[\protect\citeauthoryear{Newman}{Newman}{2005}]{Newman:2005vv}
Newman, M. (2005).
\newblock {A measure of betweenness centrality based on random walks}.
\newblock {\em Social Networks\/}~{\em 27\/}(1), 39--54.

\bibitem[\protect\citeauthoryear{Newman}{Newman}{2018}]{Newman:2018ur}
Newman, M. (2018).
\newblock {\em {Networks}\/} (2 ed.).
\newblock Oxford University Press.

\bibitem[\protect\citeauthoryear{Newman}{Newman}{2001}]{Newman:2001kc}
Newman, M. E.~J. (2001).
\newblock {Scientific collaboration networks. II. Shortest paths, weighted
  networks, and centrality}.
\newblock {\em Physical Review E\/}~{\em 64\/}(1), 016132.

\bibitem[\protect\citeauthoryear{Opsahl, Agneessens, and Skvoretz}{Opsahl
  et~al.}{2010}]{Opsahl:2010in}
Opsahl, T., F.~Agneessens, and J.~Skvoretz (2010).
\newblock {Node centrality in weighted networks: Generalizing degree and
  shortest paths}.
\newblock {\em Social Networks\/}~{\em 32\/}(3), 245--251.

\bibitem[\protect\citeauthoryear{Patuelli, Reggiani, Gorman, Nijkamp, and
  Bade}{Patuelli et~al.}{2007}]{Patuelli:2007ks}
Patuelli, R., A.~Reggiani, S.~P. Gorman, P.~Nijkamp, and F.-J. Bade (2007).
\newblock {Network Analysis of Commuting Flows: A Comparative Static Approach
  to German Data}.
\newblock {\em Networks and Spatial Economics\/}~{\em 7\/}(4), 315--331.

\bibitem[\protect\citeauthoryear{Qiao, Shan, and Zhou}{Qiao
  et~al.}{2017}]{Qiao:2017js}
Qiao, T., W.~Shan, and C.~Zhou (2017).
\newblock {How to Identify the Most Powerful Node in Complex Networks? A Novel
  Entropy Centrality Approach}.
\newblock {\em Entropy\/}~{\em 19\/}(11), 614.

\bibitem[\protect\citeauthoryear{Rodrigues}{Rodrigues}{2019}]{Rodrigues:2019ff}
Rodrigues, F.~A. (2019).
\newblock {Network Centrality: An Introduction}.
\newblock In {\em A Mathematical Modeling Approach from Nonlinear Dynamics to
  Complex Systems}, pp.\  177--196. Cham: Springer International Publishing.

\bibitem[\protect\citeauthoryear{Sabidussi}{Sabidussi}{1966}]{Sabidussi:1966wp}
Sabidussi, G. (1966).
\newblock {The centrality index of a graph}.
\newblock {\em Psychometrika\/}~{\em 31\/}(4), 581--603.

\bibitem[\protect\citeauthoryear{Schroeder, Guedes, and Duarte}{Schroeder
  et~al.}{2004}]{Schroeder:2004uv}
Schroeder, J., A.~P. Guedes, and E.~P. Duarte, Jr (2004).
\newblock {Computing the Minimum Cut and Maximum Flow of Undirected Graphs}.
\newblock Technical Report RT-DINF 003/2004.

\bibitem[\protect\citeauthoryear{Sengupta and Pal}{Sengupta and
  Pal}{2009}]{Sengupta:2009wk}
Sengupta, A. and T.~K. Pal (2009).
\newblock {\em {Fuzzy Preference Ordering of Interval Numbers in Decision
  Problems}}.
\newblock Springer Science {\&} Business Media.

\bibitem[\protect\citeauthoryear{Spadon, de~Carvalho, Rodrigues-Jr, and
  Alves}{Spadon et~al.}{2019}]{Spadon:2019bv}
Spadon, G., A.~C. P. L.~F. de~Carvalho, J.~F. Rodrigues-Jr, and L.~G.~A. Alves
  (2019).
\newblock {Reconstructing commuters network using machine learningand urban
  indicators}.
\newblock Technical Report 11801.

\bibitem[\protect\citeauthoryear{Stefanini, Guerra, and Amicizia}{Stefanini
  et~al.}{2019}]{Stefanini:2019ie}
Stefanini, L., M.~L. Guerra, and B.~Amicizia (2019).
\newblock {Interval Analysis and Calculus for Interval-Valued Functions of a
  Single Variable. Part I: Partial Orders, gH-Derivative, Monotonicity}.
\newblock {\em Axioms\/}~{\em 8\/}(4), 113.

\bibitem[\protect\citeauthoryear{Stephenson and Zelen}{Stephenson and
  Zelen}{1989}]{Stephenson:1989ug}
Stephenson, K. and M.~Zelen (1989).
\newblock {Rethinking centrality: methods and examples}.
\newblock {\em Social Networks\/}~{\em 11\/}(1), 1--37.

\bibitem[\protect\citeauthoryear{Traag}{Traag}{2014}]{Traag:2014tj}
Traag, V. (2014).
\newblock {\em {Algorithms and Dynamical Models for Communities and Reputation
  in Social Networks}}.
\newblock Springer.

\bibitem[\protect\citeauthoryear{UNCTAD}{UNCTAD}{2016}]{UNCTAD:2016a}
UNCTAD (2016).
\newblock {Merchandise trade matrix -- detailed products, exports in thousands
  of United States dollars, annual}.
\newblock Available online at:
  \url{https://https://unctadstat.unctad.org/wds/ReportFolders/reportFolders.aspx?sCS_referer=&sCS_ChosenLang=en}
  (accessed: 09.09.2016).

\bibitem[\protect\citeauthoryear{Valente and Foreman}{Valente and
  Foreman}{1998}]{Valente:1998vp}
Valente, T.~W. and R.~K. Foreman (1998).
\newblock {Integration and radiality: measuring the extent of an individual's
  connectedness and reachability in a network}.
\newblock {\em Social Networks\/}~{\em 20}, 89--105.

\bibitem[\protect\citeauthoryear{Wu, He, Zhang, Chen, Sun, Liu, Zhang, and
  Poor}{Wu et~al.}{2019}]{Wu:2019mi}
Wu, M., S.~He, Y.~Zhang, J.~Chen, Y.~Sun, Y.-Y. Liu, J.~Zhang, and H.~V. Poor
  (2019).
\newblock {A tensor-based framework for studying eigenvector multicentrality in
  multilayer networks.}
\newblock {\em Proceedings of the National Academy of Sciences\/}~{\em
  116\/}(31), 15407--15413.

\bibitem[\protect\citeauthoryear{Xu, Mao, and Bai}{Xu
  et~al.}{2016}]{2016JSMTE..03.3404X}
Xu, Q., B.~Mao, and Y.~Bai (2016).
\newblock {Network structure of subway passenger flows}.
\newblock {\em arXiv.org\/}~(3), 033404--.

\bibitem[\protect\citeauthoryear{Zeng, Liu, Qin, Wang, and Yang}{Zeng
  et~al.}{2018}]{Zeng:2018jm}
Zeng, L., J.~Liu, Y.~Qin, L.~Wang, and J.~Yang (2018).
\newblock {A Passenger Flow Control Method for Subway Network Based on Network
  Controllability}.
\newblock {\em Discrete Dynamics in Nature and Society\/}~{\em 2018\/}(6),
  1--12.

\bibitem[\protect\citeauthoryear{Zhang, Shao, He, and Gao}{Zhang
  et~al.}{2020}]{Zhang:2020gi}
Zhang, Y., C.~Shao, S.~He, and J.~Gao (2020).
\newblock {Resilience centrality in complex networks}.
\newblock {\em Physical Review E\/}~{\em 101\/}(2), 022304.

\end{thebibliography}

\newpage

\appendix
\section*{Appendix A: Lexicographic Order}
\label{Appendix_A}
Given an interval-weighted network (a triplet, see Figure~\ref{chp6_tab_intervals_lexicographic_order-triplet}), consider that each of the edge intervals is described by the lower (L) and upper limit (U) and the respective quartiles $(Q_1, Q_2, Q_3)$ (assuming a uniform distribution), as shown in Figure~\ref{chp6_tab_intervals_lexicographic_order-triplet-b}, and in the table of the Figure~\ref{chp6_tab_intervals_lexicographic_order}c (table).
\begin{figure}[H]
\centering
\begin{subfigure}[b]{0.49\linewidth}
	\centering
	\begin{tikzpicture}[inner sep=0pt, minimum size=5mm, auto,
   	node_style/.style={draw,circle,line width=.1mm, font=\fontsize{10}{10}\selectfont},
   	edge_style/.style={draw=black, ultra thick, line width=.1mm}]
    \node[node_style] (v1) at (0,1.25) {$v_1$};
    \node[node_style] (v2) at (2,2.5) {$v_2$};
    \node[node_style] (v3) at (2,0) {$v_3$};
    \draw[edge_style]  (v1) edge node[above,sloped,pos=0.5,font=\fontsize{7}{8}\selectfont] {$[2,8]$} (v2);
    \draw[edge_style]  (v1) edge node[below,sloped,pos=0.5,font=\fontsize{7}{8}\selectfont] {$[1,5]$} (v3);
    \draw[edge_style]  (v2) edge node[right=0.1,pos=0.5,font=\fontsize{7}{8}\selectfont] {$[0,10]$} (v3);
	\end{tikzpicture}
	\caption{}
	\label{chp6_tab_intervals_lexicographic_order-triplet}
\end{subfigure}
\begin{subfigure}[b]{0.49\linewidth}
	\centering
	\begin{tikzpicture}
	\draw [thick]  (0.5,0) -- (5.5,0);
	\draw (1,0) [style={font=\footnotesize}] node[below=3pt] {L};
    \draw (2,0) [style={font=\footnotesize}] node[below=3pt] {$Q_1$};
    \draw (3,0) [style={font=\footnotesize}] node[below=3pt] {$Q_2$};
    \draw (4,0) [style={font=\footnotesize}] node[below=3pt] {$Q_3$};  
    \draw (5,0) [style={font=\footnotesize}] node[below=3pt] {U};   
%
    \foreach \x in {1,2,3,4,5}
      \draw (\x cm,3pt) -- (\x cm,-3pt);
	\end{tikzpicture}
	\caption{}
	\label{chp6_tab_intervals_lexicographic_order-triplet-b}
\end{subfigure}	
\par\bigskip
\centering
\begin{subfigure}[b]{0.5\linewidth}
\centering
\fontsize{8}{10}\selectfont
\renewcommand{\arraystretch}{0.75}
\begin{tabular}{M{1cm}|M{1.5cm}|M{1.5cm}|M{1.75cm}}\thickhline
\multirow{2}{*}{} & \multicolumn{3}{c}{Intervals}   \\ \cline{2-4} 
   					& $w_{12}=[2,8]$ & $w_{13}=[1,5]$ & $w_{23}=[0,10]$ \\ \thickhline 
\rowcolor{gray!20}L	& $2$		& $1$		& $0$    \\ \hline
$Q_1$ 					& $3.5$		& $2$		& $2.5$  \\ \hline
$Q_2$					& $5$		& $3$		& $5$    \\ \hline
$Q_3$ 					& $6.5$		& $4$		& $7.5$  \\ \hline
\rowcolor{gray!20}U  	& $8$		& $5$		& $10$	\\ \thickhline     
\end{tabular}
\begin{tablenotes}
      \tiny
      \item {L, U -- interval lower and upper bounds.}
      \item {$Q_1, Q_2, Q_3$ -- 1\textsuperscript{st}, 2\textsuperscript{nd} and 3\textsuperscript{rd} quartiles.}
\end{tablenotes}
\caption{}
\end{subfigure}\vspace*{-1.75mm}
\caption{Example of the values used for each interval of the IWN.}
\label{chp6_tab_intervals_lexicographic_order}
\end{figure}
Table~\ref{chp6_Appendix_Lexicographic_order} below shows the outcome of the \textit{max-flow} calculations for each of the $5^3=125$ possible combinations, following the lexicographic order, for these three intervals. We observe that the value of the ``Max Flow'' always increases when the value considered for at least one of the values increases. Therefore, at each edge, the \textit{minimum} and \textit{maximum} flows are obtained with the corresponding \textit{minimum} and \textit{maximum} values.

\begin{landscape}
\begin{table}
\caption{Lexicographic order of three intervals each with five values.}
\label{chp6_Appendix_Lexicographic_order}
    \centering
        \includegraphics[width=1.025\linewidth,clip, trim={1cm 8.25cm 2.5cm 1.75cm}]{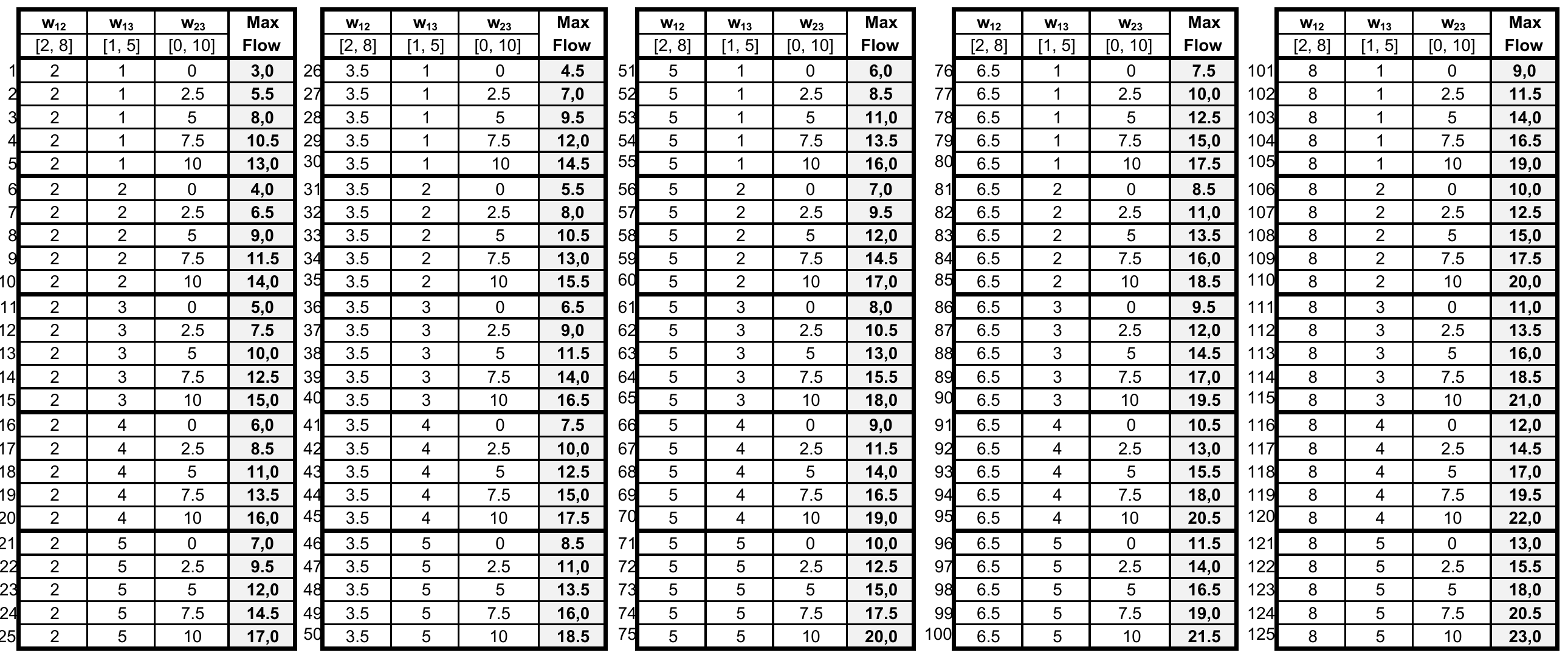}
\end{table}
\scriptsize\noindent
$w_{12},w_{13}$ and $w_{23}$ -- are interval--weighted values of the edges in a connected interval--weighted network with three vertices (triplet).\\
Max Flow -- is the Ford \& Fulkerson's \textit{maximum flow} for each of the 125 weighted networks generated in the lexicographic order.\\
\normalsize
\end{landscape}

\end{document}